\documentclass[fleqn,10pt]{wlscirep}
\usepackage[utf8]{inputenc}
\usepackage[T1]{fontenc}
\title{Reinforcement Learning Control of a Biomechanical Model of the Upper Extremity}
\usepackage{graphicx}

\author[1,*]{Florian Fischer}
\author[1]{Miroslav Bachinski}
\author[1]{Markus Klar}
\author[1]{Arthur Fleig}
\author[1]{Jörg Müller}
\affil[1]{University of Bayreuth, Bayreuth, Germany}

\affil[*]{florian.j.fischer@uni-bayreuth.de}

\usepackage{tikz, pgfplots}
\usetikzlibrary{positioning, calc}
\pgfplotsset{compat=1.12}

\usepackage{multirow}
\usepackage[caption=false]{subfig}

\usetikzlibrary{shapes,arrows,shapes.geometric}
\tikzset{%
	block/.style    = {draw, thick, rectangle}, %
	sum/.style      = {draw, circle}, %
	input/.style    = {coordinate}, %
	output/.style   = {coordinate} %
}

\newcommand{\R}{\mathbb{R}}

\definecolor{orange-red}{rgb}{1.0, 0.27, 0.0}
\definecolor{darkpastelgreen}{rgb}{0.01, 0.75, 0.24}
\definecolor{brightmaroon}{rgb}{0.76, 0.13, 0.28}

\usepackage{nicefrac}
\usepackage[normalem]{ulem} %

\DeclareMathAlphabet{\mathcal}{OMS}{cmsy}{m}{n}

\newcommand{\deletedOLD}[1]{}%

\newcommand{\deleted}[1]{}

\begin{abstract}
	Among the infinite number of possible movements that can be produced, humans are commonly assumed to choose those that optimize criteria such as minimizing movement time, subject to certain movement constraints like signal-dependent and constant motor noise. 
	While so far these assumptions have only been evaluated for simplified point-mass or planar models, we address the question of whether they can predict reaching movements in a full skeletal model of the human upper extremity. 
	We learn a control policy using a motor babbling approach as implemented in reinforcement learning, using aimed movements of the tip of the right index finger towards randomly placed 3D targets of varying size.
	We use a state-of-the-art biomechanical model, which includes seven actuated degrees of freedom.
	To deal with the curse of dimensionality, we use a simplified second-order muscle model, acting at each degree of freedom instead of individual muscles.
	The results confirm that the assumptions of signal-dependent and constant motor noise, together with the objective of movement time minimization, are sufficient for a state-of-the-art skeletal model of the human upper extremity to reproduce complex phenomena of human movement, in particular Fitts' Law and the \nicefrac{2}{3} Power Law.
	This result supports the notion that control of the complex human biomechanical system can plausibly be determined by a set of simple assumptions and can easily be learned.
\end{abstract}
\begin{document}
	
	\flushbottom
	\maketitle
	\thispagestyle{empty}

	\section*{Introduction}\label{sec:introduction}
	In the case of simple end-effector models, both Fitts' Law and the \nicefrac{2}{3} Power Law have been shown to constitute a direct consequence of minimizing movement time, under signal-dependent and constant motor noise\cite{HarrisWolpert98, Tanaka06}.
	Here, we aim to confirm that these simple assumptions are also sufficient for a full skeletal upper extremity model to reproduce these phenomena of human movement.
	As a biomechanical model of the human upper extremity, we use the skeletal structure of the {\em Upper Extremity Dynamic Model} by Saul et al.\cite{Saul14}, including thorax, right clavicle, scapula, shoulder, arm, and hand.
	The model has seven actuated degrees of freedom (DOFs): shoulder rotation, elevation and elevation plane, elbow flexion, forearm rotation, and wrist flexion and deviation.
	While the thorax is fixed in space, the right upper extremity can move freely by actuating these DOFs.
	To deal with the curse of dimensionality and make the control problem tractable, following van Beers et al.\cite{vanBeers03}, we use a simplified second-order muscle model acting at each DOF instead of individual muscles.
	These second-order dynamics map an action vector obtained from the learned policy to the resulting activations for each DOF.
	Following van Beers et al.\cite{vanBeers03}, we assume both signal-dependent and constant motor noise in the control, with noise levels 0.103 and 0.185, respectively.
	Multiplying these activations with constant moment arm scaling factors, which represent the strength of the muscle groups at the respective DOFs, yields the torques that are applied at each DOF independently.
	Further details on the biomechanical model are provided in the \textit{Methods} section below.
	
	The Upper Extremity Dynamic Model is significantly more complex than standard point-mass or linked-segment models. In particular, there is no explicit formula for the non-linear and non-deterministic system dynamics.
	Together with the objective of movement time minimization, these properties make it difficult to use classical optimal control approaches. 
	Instead, in this paper we learn a control policy using deep \textit{reinforcement learning (RL)}.
	RL algorithms, just like the optimal control methods discussed below, aim to find a policy that maximizes a given reward function.
	Moreover, they do not require any explicit knowledge about the underlying model.
	Instead, the optimal value of a certain state is estimated from sampling different actions in the environment and observing the subsequent state and obtained reward.\cite{Sutton18}

\begin{figure}[!ht]
		\centering
		\includegraphics[width=\linewidth]{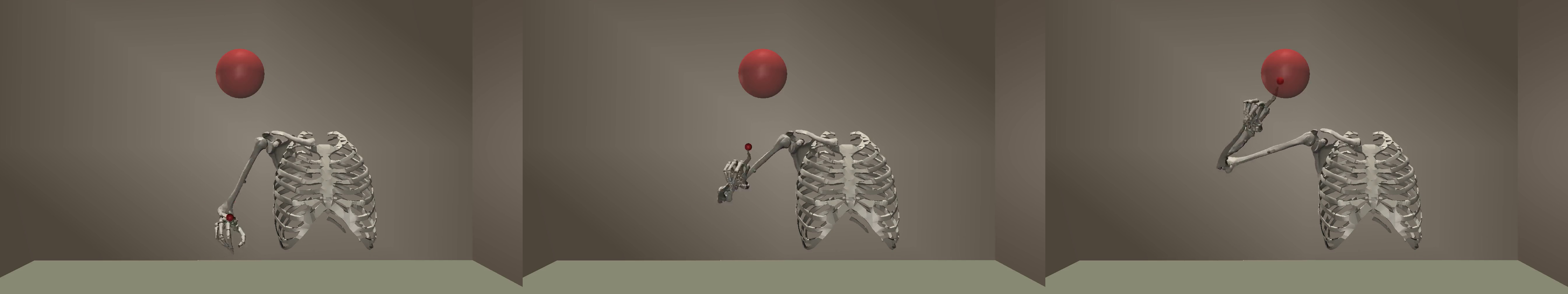}
		\caption{{\bf Synthesized reaching movement.}
			A policy implemented as a neural network computes motor control signals of simplified muscles at the joints of a biomechanical upper extremity model from observations of the current state of the upper body. We use Deep Reinforcement Learning to learn a policy that reaches random targets in minimal time, given signal-dependent and constant motor noise.
		}
		\label{fig:banner}
	\end{figure}

	In our approach, a control policy initially generates random movements, which are rewarded with the negative time to reach randomly placed 3D targets of varying size, with the right index finger (see Fig.~\ref{fig:banner}).
	This reward signal implies movement time minimization for aimed movements.
	The policy is updated using the \textit{soft-actor-critic} algorithm (SAC)\cite{Haarnoja18}.
	The actor and critic networks both consist of two fully connected layers with 256 neurons each, followed by the output layer, which either returns the means and standard deviations of the action distributions (for the actor network) or the state-action value (for the critic network).
	Further information about the network architecture and a detailed description of all state components can be found in the \textit{Methods} section below.
	To make reinforcement learning computationally feasible within a reasonable time period, a fast physics simulation is advantageous.
	Accordingly, we implemented the biomechanical model in MuJoCo\cite{Todorov12_mujoco}.
	
	It is important to note in this context that the assumption of minimizing total movement time does not provide any gradient information to the reinforcement learner.
	In particular, it is not possible to distinguish beneficial states and actions from inappropriate ones before the target has been reached, which terminates the episode and thus increases the total return. This, together with the fairly small subspace of appropriate actions relative to the number of possible control vectors, makes it very difficult to obtain a reasonable policy without additional aid.
	For this reason, we created an adaptive curriculum, which dynamically decreases the target diameter from 60 cm to less than 2 cm during training.
	This has proven to be both effective (targets with diameter around 2 cm are consistently reached by the final policy) and efficient (this minimum width was reached after 1.2M steps, while various predetermined curricula required more than 3M steps).

    \section*{Related Work}
    The question of how human arm movements are internally planned and controlled has received significant attention in the literature. 
    Important phenomena that emerge from human arm movements include Fitts' Law and the \nicefrac{2}{3} Power Law.
    In this section we review related work in these areas. 
	
	\subsection*{Motor Control}
	Many models of human motor control assume that some objective function is optimized during the planning of the movement. 
    A variety of objective functions have been proposed, including minimization of either jerk\cite{Flash85, HoffArbib93}, peak acceleration\cite{Nelson83}, end-point variance\cite{HarrisWolpert98}, duration\cite{Artstein80, Tanaka06}, or torque-change\cite{Uno89}. 
	Moreover, combined objective functions have been used to model a trade-off between different objectives, e.g., between accuracy and effort\cite{Todorov98_thesis, Li04}, or jerk and movement time\cite{Hoff94}.
	Extensions have been proposed that, e.g., focus on initial gating mechanisms\cite{Bullock88} or motor synergies representing agonist and antagonist muscle groups\cite{Plamondon98}. 
        
	While most of these models imply a separation between the planning and the execution stage, the \textit{optimal feedback control} theory\cite{Todorov02, Scott04, Todorov04, Shadmehr08, Diedrichsen10} assumes that sensory signals about the controlled quantity are fed back to the controller. These observations are then directly used to compute the remaining optimal control signals, resulting in a feedback loop.
	Extensions to infinite-horizon problems\cite{Qian12}, which yield the optimal steady-state solution at the expense of neglecting transient behavior, and explicit non-linear time costs\cite{Shadmehr10, Berret16} have been proposed.

    While many early works in motor control have modeled the biomechanics as point-mass models with linear dynamics\cite{HarrisWolpert98, Todorov98_thesis} or linked-segment models\cite{HarrisWolpert98}, there is a growing interest in biomechanical models of increasing realism and fidelity.
	This is spurred by advances in biomechanical modeling\cite{Holzbaur05, Saul14, Seth19} and simulation\cite{Delp07, Seth18}.
    Biomechanical models allow control beyond the end-effector, for example on the
    level of joints\cite{Rosenbaum95, Uno89, Nakano99, Kawato93, Kawato96, Cheema20, Berret11}, or muscles\cite{Lee19, Nakada18, Si15, Fan20}.
      
    Joint-actuated models apply different optimality criteria for movement generation and coordination, minimizing, e.g., the angular accelerations with constraints\cite{Ben-itzhak2008}, angular jerk\cite{Rosenbaum95}, torque-change\cite{Uno89, Nakano99}, mechanical energy expenditure\cite{Berret2008}, a combination of absolute work and angular acceleration\cite{Berret11}, or some combination of accuracy and effort costs in the context of optimal feedback control\cite{Li04}.
	The biomechanical plant in these works is usually represented as a linked-segment model, with simplified kinematic properties.
	In particular, the shoulder joint is commonly described as a rotation-only joint, ignoring the translatory part as well as complex movements related to the scapula and clavicle.
	Some of these models also include simplified muscles with simplified biomechanical attachment\cite{Li04}.
	
    More recently, more complex, high-fidelity biomechanical musculoskeletal models have been introduced\cite{Tieck18, Kidzinski18, Lee19, Nakada18}, where the control is muscle-based.
	In these models, the computation of the control values
	is commonly based on neural networks, particularly on deep learning and reinforcement learning.
	These methods have been applied successfully to predict coordinated muscle activations for multi-joint arm\cite{Tieck18}, lower body\cite{Kidzinski18}, and full body\cite{Lee19} movements. Moreover, a combination of 20 neural networks, each designed to imitate some specific part of the sensorimotor system, has recently been used to synthesize movements for such diverse sensorimotor tasks as reaching, writing, and drawing.\cite{Nakada18}
	To make the control of muscle-based models feasible, these works apply multiple simplifications to the full biomechanical model, such as reducing or immobilising degrees of freedom\cite{Tieck18, Nakada18} or even completely locking the movement to two dimensions\cite{Kidzinski18}, ignoring tendon's elasticity\cite{Nakada18, Lee19}, limiting maximum passive forces\cite{Lee19}, ignoring the muscle activation dynamics\cite{Nakada18, Lee19} or significantly reducing the number of independently controlled muscles\cite{Tieck18, Kidzinski18}.
	Also, the control learning strategies differ from the pure reinforcement learning approach by applying imitation learning\cite{Nakada18, Lee19}, or using artificial training data with simplified dynamics\cite{Nakada18}.
	
	Up to now, these models have not been evaluated regarding the realism of the movements generated, in particular whether they exhibit features characteristic of human movements, such as \textit{Fitts' Law} and the \textit{\nicefrac{2}{3} Power Law}.\cite{Lee19, Nakada18, Cheema20}
	
	\subsection*{Fitts' Law}

    Fitts' Law\cite{Fitts54} describes the speed-accuracy trade-off of aimed movements towards a spatially defined target. 
	Given the distance $D$ between the initial position of the controlled end-effector (e.g., the hand or the finger) and the desired target position, and given the width $W$ of the target, this law claims a logarithmic relationship between the distance-width ratio $\nicefrac{D}{W}$ and the resulting movement time $MT$:
	\begin{eqnarray}\label{eq:fitts-law-shannon}
		MT = a + b \log_{2}\left(\frac{D}{W} + 1\right),
	\end{eqnarray}
	where we used the \textit{Shannon formulation}\cite{MacKenzie89}.
	Most works that explored possible explanations for the emergence of Fitts' Law have postulated that it results from noise in motor control.
    Crossman and Goodeve\cite{CrossmanGoodeve83} showed that Fitts' Law emerges from the assumptions of isochronal submovements towards the target and constant error-velocity ratio.
    Meyer et al.\cite{Meyer88} demonstrated that a power form of Fitts' Law emerges from the optimization of the relative duration of two submovements in order to achieve minimal movement time, assuming that the standard deviation of submovement endpoints increases proportionally with movement velocity.
    Fitts' Law has also been derived within the infinite-horizon optimal control framework, assuming that the target is reached as soon as the positional end-effector variance relative to the target center falls below the desired target accuracy.\cite{Qian12}
    
	Harris and Wolpert\cite{HarrisWolpert98} proposed that the central nervous system (CNS) aims to minimize movement end-point variance given a fixed movement time, under the constraint of signal-dependent noise.
	This signal-dependent noise is assumed to be the main factor determining the end-point accuracy:
	Faster movements can be achieved through applying larger control signals (in the extreme, this leads to the time-minimizing \textit{Bang-bang} control), but only at the costs of larger deviations, which in turn induce a larger end-point variance and thus a greater risk of missing the target.
	This trade-off has a strong neuroscientific evidence\cite{Matthews96} and is consistent with the speed-accuracy trade-off proposed by Fitts' Law\cite{HarrisWolpert98, Tanaka06}.
	Moreover, in the case of arm-reaching movements, it has been shown recently that the assumptions of feed-forward control and signal-dependent noise (using dynamics of a two-link planar arm model) directly imply Fitts' Law, with coefficients $a$ and $b$ related to the level of signal-dependent noise.\cite{Takeda19}
	Both coefficients were also shown to depend on the dynamics and kinematics, e.g., on the viscosity, or the Jacobian matrix relating the joint space and the end-effector space.
	
	\subsection*{\nicefrac{2}{3} Power Law}
	Continuous, rhythmic movements such as drawing or hand-writing, exhibit a speed-curvature trade-off described by the \nicefrac{2}{3} Power Law\cite{Lacquaniti83}.
    This law proposes a non-linear relationship between the radius of curvature $\rho_{n}$ and the corresponding tangential velocity $v_{n}$, 
	\begin{eqnarray}
		v_{n} &{=}& k \rho_{n}^{1-\beta}, \\ \beta &{\approx}& \nicefrac{2}{3},
	\end{eqnarray}
	where the parameter $k$ determines the velocity gain.
	This particularly implies that higher curvature leads to lower velocity.
    It has also been demonstrated that the \nicefrac{2}{3} Power Law is equivalent to constant affine velocity\cite{Pollick97}.

	The \nicefrac{2}{3} Power Law has been confirmed to hold for a variety of task conditions, including hand movement\cite{VivianiSchneider91}, eye movement\cite{deSperati97}, perceptuomotor tasks\cite{VivianiMonoud90, Viviani97}, and locomotion\cite{Hicheur05}. Moreover, it has been shown to apply under the assumption of signal-dependent noise.\cite{HarrisWolpert98}
	Schaal and Sternad\cite{Schaal01} observed that the perimeter of the ellipse has a large impact on the validity of this law, with $\beta$ obtained from a non-linear regression
	showing deviations in the order of $30-40\%$ for large ellipses (or, alternatively, with decreasing coefficient of determination~$R^{2}$, i.e., decreasing reliability of the parameter fitting).
	Based on these observations, Schaal and Sternad argue that the \nicefrac{2}{3} Power Law cannot be an intrinsic part of the movement planning procedure, but rather occurs as a "by-product" from the generation of smooth trajectories in intrinsic joint space.\cite{Schaal01}
	Following this argumentation, the non-linearities arising from the transformation from joint space to end-effector space, i.e., from a non-trivial kinematic chain, may account for scale- and direction-dependent results.
	Other theories see the cause of the wide applicability of the \nicefrac{2}{3} Power Law either in trajectory planning processes such as motor imagery\cite{Karklinsky15} or jerk minimization\cite{Todorov98}, or directly emerging from muscle properties\cite{Gribble96} or population vectors in motor cortical areas in the CNS\cite{Schwartz94, Flash07}.
	
	\section*{Results}
	
	\subsection*{Fitts' Law}
\begin{figure}[!ht]
		\centering
		\begin{tikzpicture}
			\node (img1) at (0,0) {\includegraphics[width=\linewidth]{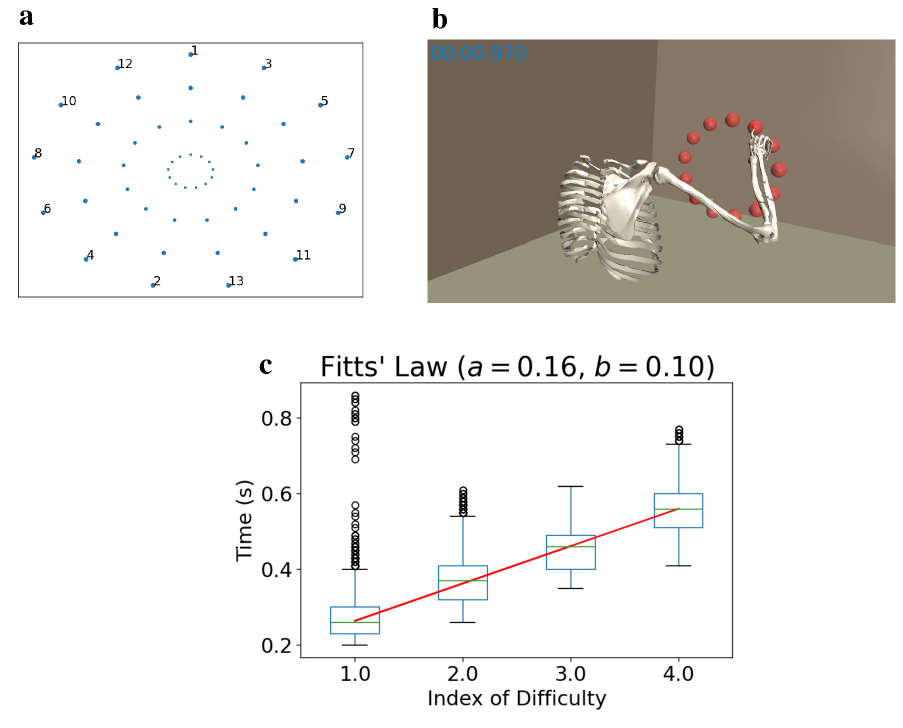}};
		\end{tikzpicture}
		
		\caption{\textbf{Fitts' Law type task.} 
			\textbf{a.} The target setup in the discrete Fitts' Law type task follows the ISO 9241-9 ergonomics standard. 
			Different circles correspond to different IDs and distances between targets.\\
			\textbf{b.} Visualization of our biomechanical model performing aimed movements. Note that for each time step, only the \textit{current} target (position and radius) is given to the learned policy. \\
			\textbf{c.} The movements generated by our learned policy conform to Fitts' Law.
			Here, movement time is plotted against ID for all distances and IDs in the considered ISO task (6500 movements in total).
		}\label{fig:fitts-law-type-task}
	\end{figure}

	In order to evaluate the trajectories resulting from our final policy for different target conditions, we designed a discrete Fitts' Law type task.
	The task follows the ISO 9241-9 ergonomics standard and incorporates 13 equidistant targets arranged in a circle at 50 cm distance in front of the body and placed 10 cm to the right of the right shoulder (Fig.~\ref{fig:fitts-law-type-task}).
	The objective is for the end-effector to reach each target and to remain inside the target for 100 ms.
	In this case we deem the movement successful.
	Although not included in the training phase, remaining inside the target seemed to be unproblematic during evaluation.
	If either the movement was successful, or 1.5 seconds have passed, the next target is given to the learned policy.

	The Index of Difficulty (ID) of the tasks ranges from 1 to 4, where ID is computed as $\log_2(D/W+1)$.
	$D$ denotes the distance between the initial and target position, and $W$ is the target size. 
	We execute 50 movements for each task condition and each direction, i.e., 6500 movements in total -- all were successful.

	Using the trajectories from this discrete pointing task, we evaluate whether the synthesized movements follow Fitts' Law\cite{Fitts54}, i.e., whether there is a linear relationship between task difficulty (ID) and the required movement time.
	Figure~\ref{fig:fitts-law-type-task}c shows the total duration for each movement sorted by ID.
	The median movement times for each ID (green lines) are approximated by a linear function (red line, with coefficient of determination $R^{2}=0.9986$).

	\subsection*{\nicefrac{2}{3} Power Law}
\begin{figure}[!ht]
		\centering
		\begin{tikzpicture}
			\node (img1) at (0,0) {\includegraphics[width=\linewidth]{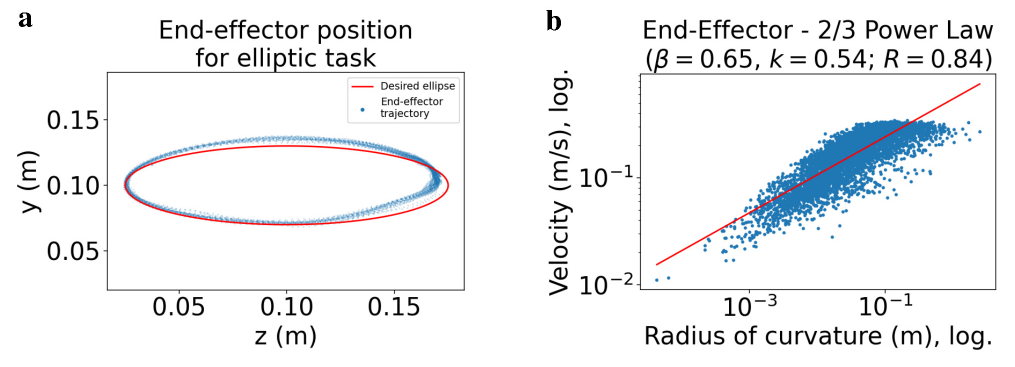}};
		\end{tikzpicture}
		\caption{{\bf Elliptic via-point task.} 
			Elliptic movements generated by our learned policy conform to the \nicefrac{2}{3} Power Law. \\
			\textbf{a.} End-effector positions projected onto the 2D space (blue dots), where targets were subsequently placed along an ellipse of 15 cm width and 6 cm height (red curve).\\
			\textbf{b.} Log-log regression of velocity against radius of curvature for end-effector positions sampled with 100 Hz when tracing the ellipse for 60 seconds.}\label{fig:powerlaw}
	\end{figure}

	We evaluate whether our model exhibits the \nicefrac{2}{3} Power Law using an elliptic via-point task.
	To this end, we define an ellipse in 2D space (55 cm in front, 10 cm above, and 10 cm to the right of the shoulder) that lies completely within the area used for target sampling during training (ellipse radii are 7.5 cm (horizontal) and 3 cm (vertical)).
	Using the via-point method described in the \textit{Methods} section below, our learned policy needs to trace the ellipse for 60 seconds as fast as possible.
	As shown in Fig.~\ref{fig:powerlaw}a), the simulation trajectories deviate from the desired ellipse, with the lower-right segment being flattened.
	Using these trajectories, we compute $\rho_{n}$ and $v_{n}$ for all time steps sampled at a rate of 100 Hz and then perform a log-log regression on the resulting values.
	This yields the optimal parameter values $\beta=0.65$ and $k=0.54$~(with correlation coefficient $R=0.84$).%
	Thus, the \nicefrac{2}{3} Power Law accounts for $71\%$ of the variance observed in elliptic movements ($R^{2}=0.71$).
	Both the data points and the linear approximation in log-log space are shown in Fig.~\ref{fig:powerlaw}b.

	\subsection*{Movement trajectories}
\begin{figure}[!ht]
		\centering
		\begin{tikzpicture}
			\node (img1) at (0,0) {\includegraphics[width=\linewidth]{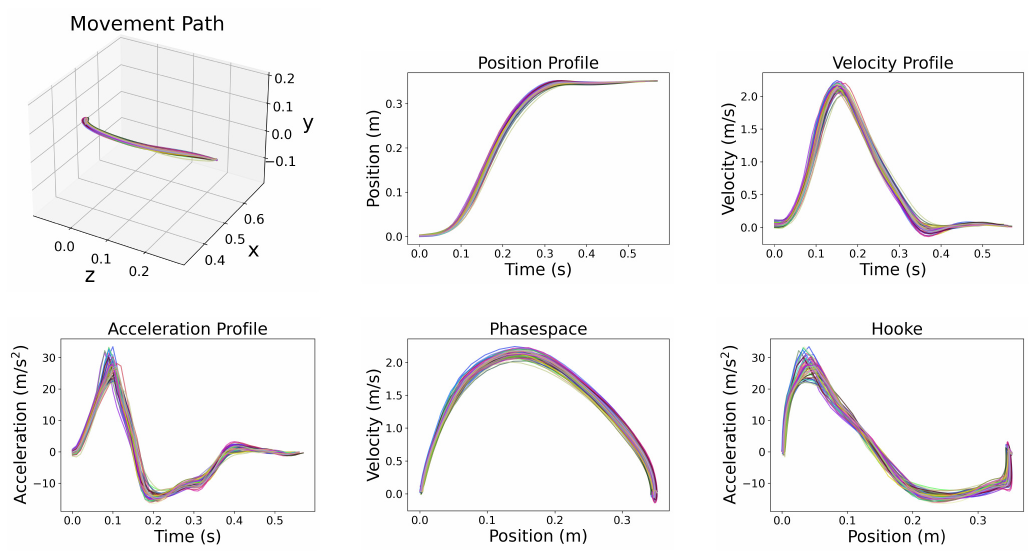}};
		\end{tikzpicture}
		
		\caption{\textbf{End-effector trajectories (ID 4).} 3D path, projected position, velocity, acceleration, phasespace, and Hooke plots of 50 aimed movements (between targets 7 and 8 shown in Fig.~\ref{fig:fitts-law-type-task}a) with ID 4 and a target distance of 35 cm.
		}\label{fig:endeffector-ID4}
	\end{figure}

\begin{figure}[!ht]
		\centering
		\begin{tikzpicture}
			\node (img1) at (0,0) {\includegraphics[width=\linewidth]{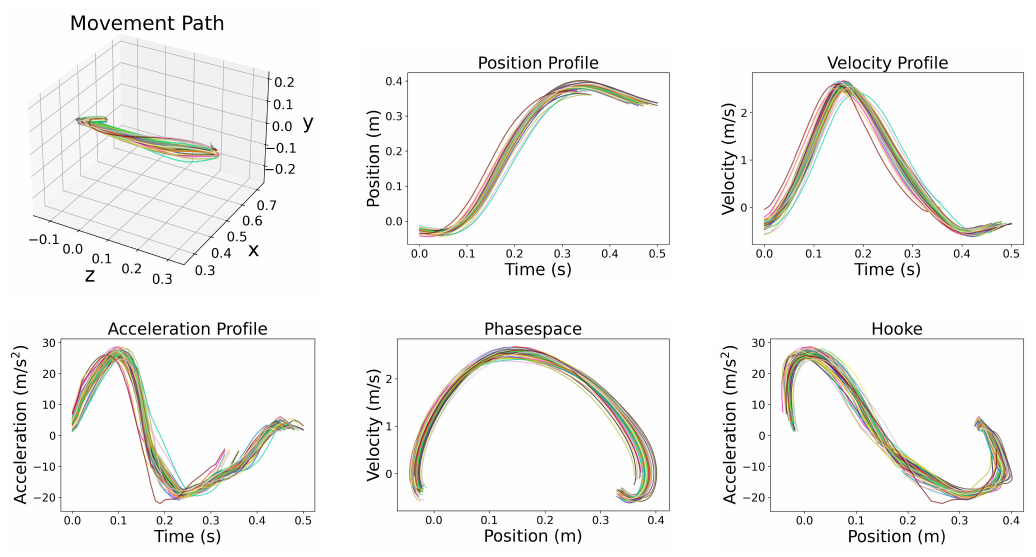}};
		\end{tikzpicture}
		
		\caption{\textbf{End-effector trajectories (ID 2).} 3D path, projected position, velocity, acceleration, phasespace, and Hooke plots of 50 aimed movements (between targets 7 and 8 shown in Fig.~\ref{fig:fitts-law-type-task}a) with ID 2 and a target distance of 35 cm.}\label{fig:endeffector-ID2}
	\end{figure}

	In addition to Fitts' Law and the \nicefrac{2}{3} Power Law, we qualitatively analyze the movement trajectories generated by the model.
	Figures~\ref{fig:endeffector-ID4} and~\ref{fig:endeffector-ID2} show the position, velocity, and acceleration time series, as well as 3D movement path, Phasespace, and Hooke plots for multiple movements from the Fitts' Law type task for two representative task conditions (ID~4 respective ID~2, each with a 35 cm distance between targets) and one representative movement direction (between targets 7 and 8 shown in Fig.~\ref{fig:fitts-law-type-task}a).
	Apart from the 3D movement path, all plots show centroid projections of the respective trajectory onto the vector between the initial and target positions.
	
	The movements exhibit typical features of human aimed movements, such as symmetric bell-shaped velocity profiles.\cite{Morasso81}
	Movements are smooth, and gently accelerate and decelerate, as evident in the acceleration profiles and Hooke plots in Fig.~\ref{fig:endeffector-ID4} and Fig.~\ref{fig:endeffector-ID2}.
	For high ID (Fig.~\ref{fig:endeffector-ID4}), movements exhibit an initial rapid movement towards the target, followed by an extended phase of corrective movements.
	For low ID (Fig.~\ref{fig:endeffector-ID2}), the phase of corrective movements is generally shorter.
	
	Movement trajectories towards the target are slightly curved and some of them exhibit pronounced correctional submovements at the end (see, e.g., Supplementary Fig. S1 and S2 online). 
	The between-movement variability within one movement direction and task condition decreases with increasing ID. In particular, very simple ID 1 movements exhibit a large variability and are most prone to outliers (see, e.g., Supplementary Fig. S3 online).
	
	For a few movement directions (mostly in ID 2 tasks), the corresponding plots seem to incorporate two different  trajectory types (see, e.g., Supplementary Fig. S6 online): 
	While some movements start with zero or even a negative acceleration and show a typical N-shaped acceleration profile, others exhibit a positive acceleration at the beginning, and their first peak is less pronounced. 
	The reason for this behavior is corrective submovements at the end of the previous movement (see, e.g., Supplementary Fig. S4 and S5 online), leading to a different initial acceleration at the beginning of the subsequent movement.    
	Apart from these notable features, almost all movements exhibit bell-shaped velocity and N-shaped acceleration profiles, as is typical for pointing tasks\cite{Morasso81, Abend82}.

	\section*{Discussion}
	Our results indicate that, under the assumption of movement time minimization given signal-dependent and constant motor noise, movement of the human upper extremity model produced by reinforcement learning follows both Fitts' Law and the \nicefrac{2}{3} Power Law. 
    The movement times of aimed movements produced by the model depend linearly on the Index of Difficulty of the movement.
    For elliptic movements, the logarithm of the velocity of the end-effector correlates with the logarithm of the radius of curvature.\cite{Cohen13}
	The optimal $\beta=0.65$ obtained from log-log regression matches the proposed value of \nicefrac{2}{3}, with a correlation coefficient of $R=0.84$, which is consistent with previous observations, as the required ellipse has moderate size\cite{Schaal01}.
	Finally, the generated trajectories exhibit features that are typical for human arm movements, such as bell-shaped velocity and N-shaped acceleration profiles. 

    The results confirm previous findings that demonstrated these phenomena in simpler models of the human biomechanics. 
    In particular, the emergence of Fitts' Law and the \nicefrac{2}{3} Power Law from the assumption of signal-dependent noise has been demonstrated in the case of point-mass and linked-segment models of the human arm.\cite{HarrisWolpert98, Tanaka06, Takeda19} 
    Our results support that insight by showing that Fitts' Law and the \nicefrac{2}{3} Power Law also emerge from those normative principles in a state-of-the-art biomechanical model of the human arm with simplified actuation.
          
	In addition, we want to emphasize that the control signals that drive this model were obtained from RL methods. 
	The fact that Fitts' Law and the \nicefrac{2}{3} Power Law hold for the generated trajectories 
    provides evidence that behavior abiding these established laws of human motion can be generated using joint-actuated biomechanical models controlled by reinforcement learning algorithms.
	To the best of our knowledge, this has not yet been shown for state-of-the-art biomechanical models.

	One limitation of our approach is the implicit assumption of \textit{perfect observability}, as all state information (joint angles, end-effector position, etc.) are immediately available to the controller, without any disturbing noise.
	In the future, it will be interesting to combine state-of-the art models of sensory perception with the presented RL-based motor control approach.
	Promising approaches to address this problem include the usage of recurrent networks\cite{Hausknecht15, Liu20} and the internal formation of ``beliefs'', given the latest (imperfect) observations\cite{Igl18}.
	
	Another limitation is the usage of simplified muscle dynamics due to the curse of dimensionality. However, recent applications of deep learning methods, which approximate complex state-dependent torque limits\cite{Jiang19} or muscle activation signals\cite{Nakada18} using synthesized training data, raise hope for future approaches that efficiently combine RL or optimal control methods with state-of-the-art muscle models.
	It will be interesting to see whether well-established regularities such as Fitts' Law or the \nicefrac{2}{3} Power Law also emerge from such models.

	\section*{Methods}\label{sec:methods}
	
	Below, we first provide details on our biomechanical model.
	After discussing our general reinforcement learning approach, we focus on the individual components of our method, namely states, actions, scaling factors, rewards, and an adaptive target-selection mechanism.
	We also provide details on the implementation of our algorithm.
	Finally, we discuss the methods used for evaluation.
	
	\subsection*{Biomechanical Model of the Human Upper Extremity}
	Our biomechanical model of the human upper extremity is based on the {\em Upper Extremity Dynamic model}\cite{Saul14}, which was originally implemented in \textit{OpenSim}\cite{Delp07}. 
	Kinematically, the model represents the human shoulder and arm, using seven physical bodies and five ''phantom'' bodies to model the complex movement of the shoulder.
	This corresponds to three joints (shoulder, elbow, and wrist) with seven DOFs and five additional joints with thirteen associated components coupled by thirteen constraints with the DOFs.
	Each DOF has constrained joint ranges (see Table~\ref{tab:joint_ranges}), which limits the possible movements.
	In contrast to linked-segment models, the \textit{Upper Extremity Dynamic model} represents both translational and rotatory components of the movement within shoulder, clavicle, and scapula, and also within the wrist. 
	It also contains physiological joint axis orientations instead of the perpendicular orientations in linked-segment models.
	The dynamics components of the musculoskeletal model are represented by the weight and inertia matrix of each non-phantom body and the default negligible masses and inertia of all phantom bodies. 
	The dynamics properties of the model were extracted from various previously published works on human and cadaveric studies.
	The active components of the {\em Upper Extremity Dynamic Model} consist of thirty-one Hill-type muscles as well as of fourteen coordinate limit forces softly generated by the ligaments when a DOF approaches the angle range limit.
	Further details of this model are given in Saul et al.\cite{Saul14}

	In order to make reinforcement learning feasible, we manually implement the {\em Upper Extremity Dynamic Model} in the fast MuJoCo physics simulation\cite{Todorov12_mujoco}. 
	With respect to kinematics, the MuJoCo implementation of the model is equivalent to the original OpenSim model and contains physiologically accurate degrees of freedom, as well as corresponding constraints.
	We assume the same physiological masses and inertial properties of individual segments as in the OpenSim model. 
	We do not implement muscles in the MuJoCo model, as this would significantly slow down the simulation and make reinforcement learning computationally infeasible due to the exponential growth of decision variables in the (discretized) action space
	when increasing the number of DOFs -- the \textit{curse of dimensionality}.
	In particular, computing dynamic actuator lengths (which significantly affect the forces produced by muscle activation patterns) has proven challenging in MuJoCo\cite{Ikkala20}.
	Instead, we implement simplified actuators, representing aggregated muscle actions on each individual DOF, which are controlled using the second-order dynamics
	introduced by van der Helm et al.\cite{vanderHelm00} with fixed excitation and activation time constants $t_{e}=30$ ms and $t_{a}=40$ ms, respectively.
	We discretize the continuous state space system using the \textit{forward Euler method}, which yields the following dynamics: 
	\begin{eqnarray}\label{eq:activationmodel}
		\begin{bmatrix}
			\sigma_{n+1}^{(q)} \\ 
			\dot{\sigma}_{n+1}^{(q)}
		\end{bmatrix} =
		\begin{bmatrix}
			1 & \Delta t \\
			\frac{-\Delta t}{(t_e t_a)} & 1 - \Delta t \frac{t_e + t_a}{t_e t_a}
		\end{bmatrix} 
		\begin{bmatrix}
			\sigma_{n}^{(q)} \\ 
			\dot{\sigma}_{n}^{(q)}
		\end{bmatrix}
		+ \begin{bmatrix}
			0 \\ \frac{\Delta t}{t_e t_a}
		\end{bmatrix} c_{n}^{(q)},
	\end{eqnarray}
	where $c_{n}^{(q)}$ is the applied control and $\sigma_{n}^{(q)}$ the resulting activation for each DOF $q\in\mathcal{Q}$, and $\mathcal{Q}$ is the set that contains all DOFs.
	The controls are updated every $\Delta t$=10 ms, at time steps $n\in\{0, \dots, N-1\}$. 
	To get more accurate results, at each time step $n$, we compute five sub-steps (during which the control $c_{n}^{(q)}$ is constant) with a sampling time of 2~ms to arrive at time step $n+1$.
		
	We assume both signal-dependent and constant noise in the control, i.e.,
	\begin{eqnarray}\label{eq:noisemodel}
		c_{n}^{(q)} = (1 + \eta_{n}) a_{n}^{(q)} + \epsilon_{n},
	\end{eqnarray}
	where $a_{n}=(a_{n}^{(q)})_{q\in\mathcal{Q}}$ denotes the action vector obtained from the learned policy, and $\eta_{n}$ and $\epsilon_{n}$ are Gaussian random variables with zero mean and standard deviations of 0.103 and 0.185, respectively, as described by van Beers et al.\cite{vanBeers03}
	The torques, which are applied at each DOF independently, are obtained by multiplying the respective activation $\sigma_{n}^{(q)}$ with a constant scaling factor $g^{(q)}$, which represents the strength of the muscle groups at the this DOF, i.e., 
	\begin{eqnarray}\label{eq:torquemodel}
		\tau_{n}^{(q)} = g^{(q)} \sigma_{n}^{(q)}.
	\end{eqnarray}
	We select the scaling factors, and respectively the maximum voluntary torques for the actuators given in Table~\ref{tab:joint_ranges}, based on experimental data as described below.
	We currently do not model the soft joint ranges in MuJoCo, as the movements the model produces do not usually reach joint limits.
	
	The used biomechanical model provides the following advantages over simple linked-segment models:
	\begin{itemize}
		\item phantom bodies and joints allow for more realistic movements, including both translation and rotation components within an individual joint,
		\item individual joint angle and torque limits are set for each and every DOF,
		\item axes between joints are chosen specifically and not just perpendicular between two segments,
		\item the model includes physiological body segment masses, and yields better options for scaling individual body parts, e.g., based on particular individuals. 
	\end{itemize}
	
	\subsection*{Reinforcement Learning}\label{sec:reinforcement-learning}
	We define the task of controlling the biomechanical model of the human upper extremity through motor control signals applied at the joints as a reinforcement learning problem, similar to recent work from Cheema et al.\cite{Cheema20}
	In this formulation, a policy $\pi_\theta(a|s)$ models the conditional distribution over actions $a \in \mathcal{A}$ (motor control signals applied at the individual DOFs) given the state $s \in \mathcal{S}$ (the pose, velocities, distance to target, etc.).
	The subindex $\theta$ denotes the parameters of the neural networks introduced below.
	At each timestep $n\in\{0, \dots, N\}$, we observe the current state $s_n$, and sample a new action $a_n$ from the current policy $\pi_\theta$.
	The physical effects of that action, i.e., the application of these motor control signals, constitute the new state $s_{n+1}$, which we obtain from our biomechanical simulation.
	In our model, given $s_{n}$ and $a_{n}$, the next state $s_{n+1}$ is not deterministic, since both signal-dependent and constant noise are included.
	Hence, we denote the probability of reaching some subsequent state $s_{n+1}$ given $s_{n}$ and $a_{n}$ by $p(s_{n+1}|s_{n}, a_{n})$, while $p(s_0)$ denotes the probability of starting in $s_0$.
	Given some policy $\pi_{\theta}$ and a trajectory $T=(s_0, a_0, \dots , a_{N-1}, s_N)$,
	\begin{eqnarray}\label{eq:probability-trajectory}
		p_\theta(T) = p(s_0)\prod_{n=0}^{N-1}\pi_\theta(a_n|s_n)p(s_{n+1}|s_{n}, a_{n})
	\end{eqnarray}
	describes the probability of realizing that trajectory. 
	Evaluating/Sampling equation~\eqref{eq:probability-trajectory} for all possible trajectories $T\in\mathcal{T}$ then yields the distribution over all possible trajectories, $\varrho_\theta^{\mathcal{T}}$, induced by a policy~$\pi_{\theta}$.
	
	We compute a reward $r_n$ at each time step $n$, which allows us to penalize the total time needed to reach a given target.
	The total return of a trajectory is given by the sum of the (discounted) rewards $\sum_{n=0}^{N}\gamma^n r_n$,
	where ${\gamma\in]0,1]}$ is a discount factor that causes the learner to prefer earlier rewards to later ones.
	Incorporating the entropy term,
	\begin{eqnarray}
		\mathcal{H}(\pi_\theta(\cdot \mid s))=\mathbb{E}_{a\sim\pi_\theta(\cdot \mid s)}[-\log(\pi_\theta(a\mid s))],
	\end{eqnarray}
 	yields the expected (soft) return
	\begin{eqnarray}\label{eq:soft-return}
		\label{eq:expected_return_function}
		J(\theta) = \mathbb{E}_{T\sim \varrho_\theta^{\mathcal{T}}} \left[\left(\sum_{n=0}^{N-1} \gamma^{n} \left(r_n - \alpha \log(\pi_{\theta}(a_{n} \mid s_{n}))\right)\right) + \gamma^{N}r_{N}\right],
	\end{eqnarray}
	which we want to maximize with respect to the parameters~$\theta$, i.e., the goal is to identify the optimal parameters $\theta^*$ that maximize $J(\theta)$.
	Here, the \textit{temperature parameter $\alpha > 0$} determines the importance of assigning the same probability to all actions that yield the same return (enforced by maximizing the entropy $\mathcal{H}$), i.e., increasing the "stochasticity" of the policy $\pi_{\theta}$, relative to maximizing the expected total return.
	It thus significantly affects the optimal policy, and finding an "appropriate" value is non-trivial and heavily depends on the magnitude of the rewards $r_{n}$. For this reason, we decided to adjust it automatically during training together with the parameters $\theta$, using dual gradient descent as implemented in the soft-actor critic algorithm (see below)~\cite{Haarnoja18}.
	
	It is important to note that the soft return in Equation~\eqref{eq:soft-return} is different from the objective function used in standard reinforcement learning. The MaxEnt RL formulation, which incorporates an additional entropy maximization term, provides several technical advantages. These include the natural state-space exploration~\cite{Mnih16, Eysenbach18}, a smoother optimization landscape that eases convergence towards the global optimum~\cite{Ahmed19, Fox15, Vieillard21}, and increased robustness to changes in the reward function~\cite{Eysenbach19, Eysenbach21}. In practice, many RL algorithms have gained increased stability from the additional entropy maximization~\cite{Haarnoja18b, Abdolmaleki18, Levine18}. Conceptually, MaxEnt RL can be considered equivalent to \textit{probabilistic matching}, which has been used to explain human decision making~\cite{Vulkan00, Gruenwald04}. Existing evidence indicates that human adults tend to apply probabilistic matching methods rather than pure maximization strategies~\cite{Weir64, Gallistel90, Vulkan00}. However, these observations still lack conclusive neuroscientific explanation~\cite{Abdolmaleki18}.

	In order to approximate the optimal parameters $\theta^*$, we use a policy-gradient approach, which iteratively refines the parameters $\theta$ in the direction of increasing rewards.
	Reinforcement learning methods that are based on fully sampled trajectories usually suffer from updates with high variance.
	To reduce this variance and thus accelerate the learning process, we choose an approach that includes two approximators: an \textit{actor network} and a \textit{critic network}.
	These work as follows.
	Given some state $s_0$ as input, the actor network outputs the (standardized) mean and standard deviation of as many normal distributions as dimensions of the action space. 
	The individual action components are then sampled from these distributions.
	To update the actor network weights, we must measure the ``desirability'' of some action $a$, given some state $s$, i.e., how much reward can be expected when starting in this state with this action and subsequently following the current policy.
	These values are approximated by the critic network.%

\begin{figure}[!ht]
		\centering
		\begin{tikzpicture}
			\node (img1) at (0,0) {\includegraphics[width=\linewidth]{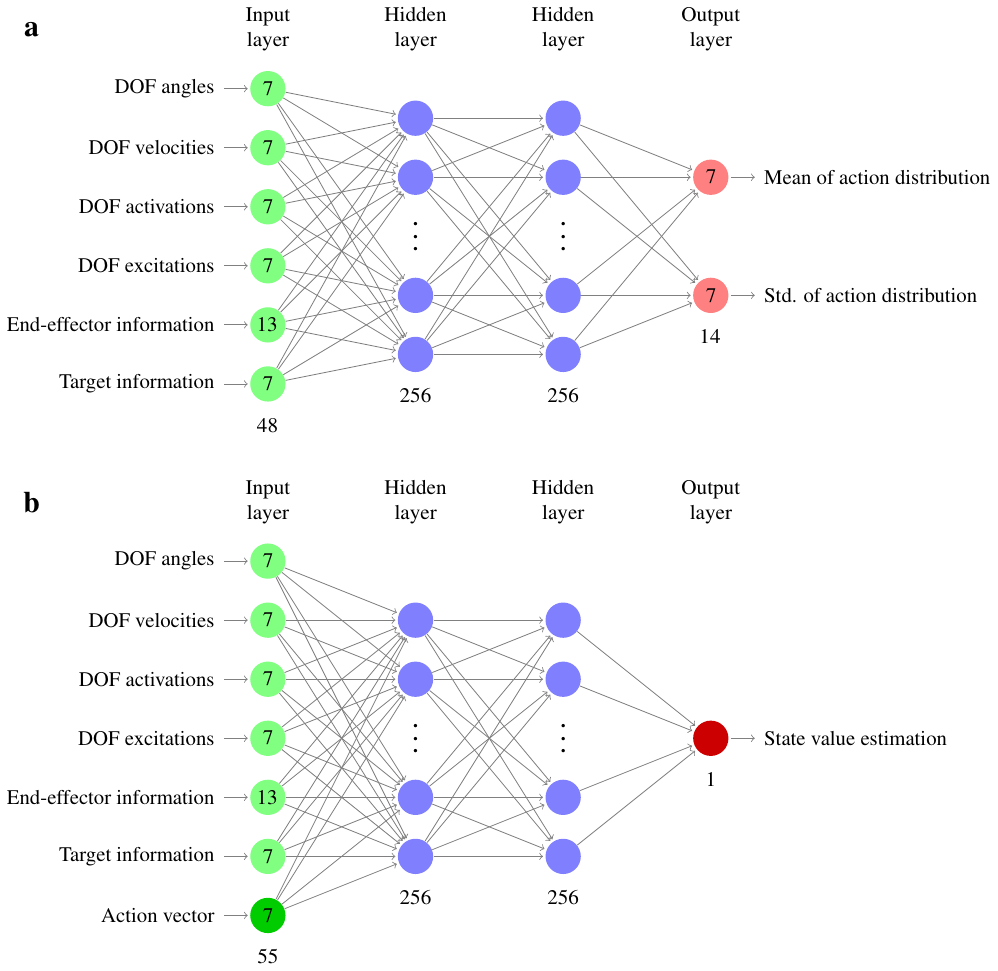}};
		\end{tikzpicture}
		\caption{\textbf{Neuronal network architectures.} 
		\textbf{a.} The actor network takes a state $s$ as input and returns the policy $\pi_{\theta}$ in terms of mean and standard deviation of the seven normal distributions, from which the components of the action vector are drawn.
		\textbf{b.} The critic network takes both state $s$ and action vector $a$ as input and returns the estimated state-action value.
		Two critic networks are trained simultaneously to improve the speed and stability of learning (\textit{Double Q-Learning}). \\
		Detailed information about the input state components are given in the \textit{Methods} section.
		}\label{fig:nn-architectures}
	\end{figure}

	The architecture of both networks is depicted in Fig.~\ref{fig:nn-architectures}. 
	For the sake of a simpler notation, the parameter vector~$\theta$ contains the weights of both networks, however these weights are not shared between the two.
	These two networks
	are then coupled with the \textit{soft actor-critic (SAC)} algorithm\cite{Haarnoja18}, which has been used successfully in physics-based character motion\cite{Peng18}:
	As a policy-gradient method, it can be easily used with a continuous action space such as continuous motor signals -- something that is not directly possible with value function methods like \textit{DQN}\cite{Sutton18}.
	As an off-policy method that makes use of a replay buffer, it is quite sample-efficient.
	This is important, since running forward physics simulations in MuJoCo constitutes the major part of the training duration.
	Moreover, it has been shown that SAC outperforms other state-of-the-art algorithms such as PPO\cite{Schulman17} or TD3\cite{Fujimoto18}. 
	Supporting the observations in Haarnoja et al.\cite{Haarnoja18}, we also found our training process to be faster and more robust when using SAC rather than PPO.
	Moreover, SAC incorporates an automatic adaption of the temperature $\alpha$ using dual gradient descent, which eliminates the need for manual, task-dependent fine-tuning.
	In order to obtain an unbiased estimate of the optimal value function, we use \textit{Double Q-Learning}\cite{Hasselt10}, using a separate target critic network.
	The neural network parameters are optimized with the Adam optimizer\cite{Kingma14}.

	\subsection*{States, Actions, and Scaling Factors}
	
	Using the MuJoCo implementation of the biomechanical model described above, the \textbf{states} $s\in\mathcal{S}\subseteq\R^{48}$ in our RL approach include the following information:
	\begin{itemize}
		\item joint angle for each DOF $q \in \mathcal{Q}$ in radians (7 values),
		\item joint velocity for each DOF $q \in \mathcal{Q}$ in radians/s (7 values),
		\item activations $\sigma^{(q)}$ and excitations $\dot{\sigma}^{(q)}$ for each DOF $q \in \mathcal{Q}$  ($2\times7$ values),
		\item positions of the end-effector and target sphere ($2\times3$ values),
		\item (positional) velocities of the end-effector and target sphere ($2\times3$ values),
		\item (positional) acceleration of the end-effector (3 values),
		\item \textit{difference vector}: vector between the end-effector attached to the index finger and the target, pointing towards the target (3 values),
		\item projection of the end-effector velocity towards the target (1 value),
		\item radius of the target sphere (1 value).
	\end{itemize}
	We found that in our case, the target velocity (which always equals zero for the considered tasks), the end-effector acceleration, the difference vector, and the projection of the end-effector velocity can be omitted from state space without reducing the quality of the resulting policy. However, we decided to incorporate these observations, as they did not considerably slow down training and might be beneficial for more complex tasks such as target tracking or via-point tasks.
	
	Each component $a^{(q)}\in\left[-1,1\right]$ of the \textbf{action vector} $a=(a^{(q)})_{q\in\mathcal{Q}}\in\mathcal{A}$ $\subseteq\R^{7}$ is used to actuate some DOF $q\in\mathcal{Q}$ by applying the torque $\tau^{(q)}$ resulting from equations~\eqref{eq:activationmodel}-\eqref{eq:torquemodel}.
	Note that in addition to these actuated forces, additional active forces (e.g., torques applied to parent joints) and passive forces (e.g., gravitational and contact forces) act on the joints in each time step.
	
	We determine experimentally the maximum torque a human would exert at each DOF in this task as follows.
	We implemented the Fitts' Law task described above in a VR environment displayed via the HTC Vive Pro VR headset.
	We recorded the movements of a single participant performing the task, using the Phasespace X2E motion capture system
	with a full-body suit provided with 14 optical markers.
	This study was granted ethical approval by the ethics committee of the University of Bayreuth and followed ethical standards according to the Helsinki Declaration. Written informed consent was received from the participant, which received an economic compensation for participating in the study.
	Using OpenSim, we scaled the {\em Upper Extremity Dynamic Model} to this particular person.
	We then used OpenSim to perform \textit{Inverse Dynamics} to obtain the torque sequences that are most likely to produce the recorded marker trajectories.
	For each DOF $q \in \mathcal{Q}$, we set the corresponding \textbf{scaling factor} $g^{(q)}$ to the absolute maximum torque applied at this DOF during the experiment, omitting a small number of outliers from the set of torques, i.e., values with a distance to mean larger than $20$ times the standard deviation.
	The resulting values are shown in Table~\ref{tab:joint_ranges}.
	
	\subsection*{Reward Function and Curriculum Learning}
	The behavior of the policy is determined largely by the reward $r_n$ that appears in equation~\eqref{eq:expected_return_function}. 
	We designed the reward following Harris and Wolpert\cite{HarrisWolpert98}, who argue that there is no rational explanation as to why the central nervous system (CNS) should explicitly try to minimize previously proposed metrics such as the change in torque applied at the joints\cite{Uno89}, or the acceleration (or jerk) of the end-effector\cite{Flash85}. 
	They argue that it is not even clear whether the CNS is able to compute, store, and integrate these quantities while executing motions.
	
	Instead, they argue that the CNS aims to minimize movement end-point variance given a fixed movement time, under the constraint of signal-dependent noise.
	Following Harris and Wolpert\cite{HarrisWolpert98}, this is equivalent to minimizing movement time when the permissible end-point variance is given by the size of the target.
	This objective is simple and intuitively plausible, since achieving accurate aimed movements in minimal time is critical for the success of many movement tasks.
	Moreover, it has already been applied to linear dynamics\cite{Tanaka06}.
	
	Therefore, the objective of our model is to \textit{minimize movement time while reaching a target of given width}.
	
	More precisely, our \textbf{reward function} consists only of a time reward, which penalizes every time step of an episode equally:
	\begin{eqnarray}\label{eq:reward_time}
		r_n = - 100 \Delta t.
	\end{eqnarray}
	This term provides incentives to terminate the episode (which can only be achieved by reaching the target) as early as possible.
	Since we apply each control $a_{n}$ for 10 ms, $\Delta t$ amounts to $0.01$ in our case, i.e., $r_{n}=-1$ in each time step $n\in\{0,\dots,N\}$.

	According to our experience, it is possible to learn aimed movements despite the lack of gradient provided by the reward function, provided the following requirements are met.
	The initial posture needs to be sampled randomly, and the targets need to be large enough at the beginning of the training to ensure that the target is reached by exploration sufficiently often in early training steps to guide the reinforcement learner.
	However, creating a predetermined curriculum that gradually decreases the target width during training appropriately has proved very difficult. 
	In most cases, the task difficulty either increased too fast, leading to unnatural movements that do not reach the target directly (and often not at all), or progress was slow, resulting in a time-consuming training phase.
	
	For this reason, we decided to use an adaptive curriculum, which adjusts the target width dynamically, depending on the recent success rate.
	Specifically, we define a \textit{curriculum state}, which is initialized with an initial target diameter of 60 cm. Every 10K update steps, the current policy is evaluated on 30 complete episodes, for which target diameters are chosen, depending on the current state of the curriculum. Based on the percentage of targets reached within the permitted 1.5 seconds (\textit{success rate}), the curriculum state is updated.
	If the success rate falls below $70\%$, it is increased by 1~cm; if the success rate exceeds $90\%$, it is decreased by 1~cm.
	To avoid target sizes that are larger than the initial width or are too close to zero, we clipped the resulting value to the interval $\left[0.1~\text{cm},~60~\text{cm}\right]$.
	
	At the beginning of each episode, the target diameter is set to the current curriculum state with probability $1 - \varepsilon$, and sampled uniformly randomly between 0.1 cm and 60 cm with probability $\varepsilon=0.1$, which has proven to be a reasonable choice.
	This ensures in particular that all required target sizes occur throughout the training phase, and thus prevents forgetting how to solve ``simpler'' tasks (in literature, often referred to as \textit{catastrophic forgetting}; see, e.g., McCloskey et al.\cite{McCloskey89}).
	
	\subsection*{Implementation of the Reinforcement Learning Algorithm}
	
	The actor and critic networks described in the \textit{Reinforcement Learning} section consist of two fully connected layers with 256 neurons each, followed by the output layer, which either returns the means and standard deviations of the action distributions (for the actor network) or the state-action value (for the critic network).
	To improve the speed and stability of learning, we train two separate, but identically structured critic networks and use the minimum of both outputs as the teaching signal for all networks (\textit{Double Q-Learning})\cite{Hasselt10, Haarnoja18}.
	In all networks, ReLU\cite{Nair10} is used as non-linearity for both hidden layers.
	The network architectures are depicted in Fig.~\ref{fig:nn-architectures}.
	
	The reinforcement learning methods of our implementation are based on the \textit{TF-Agents} library\cite{TFAgents}.
	The learning phase consists of two parts, which are repeated alternately: \textit{trajectory sampling} and \textit{policy updating}.
	
	In the trajectory sampling part, the target position is sampled from the uniform distribution on a cuboid of 70 cm height, 40 cm width, and 30 cm depth, whose center is placed 50 cm in front of the human body, and 10 cm to the right of the shoulder.
	The width of the target is controlled by the adaptive curriculum described above.
	The biomechanical model is initialized with some random posture, for which the joint angles are uniformly sampled from the convex hull of static postures that enables keeping the end-effector in one of 12 targets placed along the vertices of the cuboid described above. The initial joint velocities are uniformly sampled from the interval $\left[-0.005~\text{radians/s},~0.005~\text{radians/s}\right]$.
	
	In each step $n\in\{0, \dots, N-1\}$, given the current state vector $s_{n}\in\mathcal{S}$ (see description above), an action is sampled from the current policy $\pi_{\theta}(\cdot \mid s_{n})$.
	Next, the MuJoCo simulation uses this action to actuate the model joints. It also updates the body posture, and returns both the reward $r_n$ and the subsequent state vector $s_{n+1}$.
	In our implementation, each episode in the learning process contains at most $N=150$ of such steps, with each step corresponding to 10 ms (allowing movements to be longer than one and a half seconds did not improve the training procedure significantly).
	If the target is reached earlier, i.e., the distance between end-effector and target center is lower than the radius of the target sphere, and the end-effector remained inside the target for $100$ ms, the current episode terminates and the next episode begins with a new target position and width.
	At the beginning of the training, 10K steps are taken and the corresponding transitions stored in a replay buffer, which has a capacity of 1M steps.
	During training, only one step is taken and stored per sampling phase.
	
	In the policy updating part, 256 previously sampled transitions $(s_{n}, a_{n}, r_{n}, s_{n+1})$ are randomly chosen from the replay buffer to update both the actor network and the critic network weights.
	We use a discount factor of $\gamma=0.99$ in the critic loss function of SAC.
	All other parameters are set to the default values of the TF-Agents SAC implementation\cite{TFAgents}.
	
	Both parts of our learning algorithm, the trajectory sampling and the policy update, are executed alternately until the curriculum state, i.e., the current suggested target diameter, falls below 1 cm.
	With our implementation, this was the case after 1.2M steps, corresponding to about four hours of training time. 
	To evaluate a policy $\pi_\theta$, we apply the action $a_{n}^{*}$ with the highest probability under this policy for each time step (i.e., we use the corresponding \textit{greedy} policy) and evaluate the resulting trajectory. 
	Such an \textit{evaluation} is done every 10K steps, for which 30 complete episodes are generated using this deterministic policy, and the resulting performance indicators are stored. 
	After the training phase, $\theta^{*}$ is set to the latest parameter set $\theta$, i.e., the final policy $\pi_{\theta^{*}}$ is chosen as the latest policy $\pi_{\theta}$.
	
\begin{figure}[!ht]
		\centering
		\begin{tikzpicture}
			\node (img1) at (0,0) {\includegraphics[width=\linewidth]{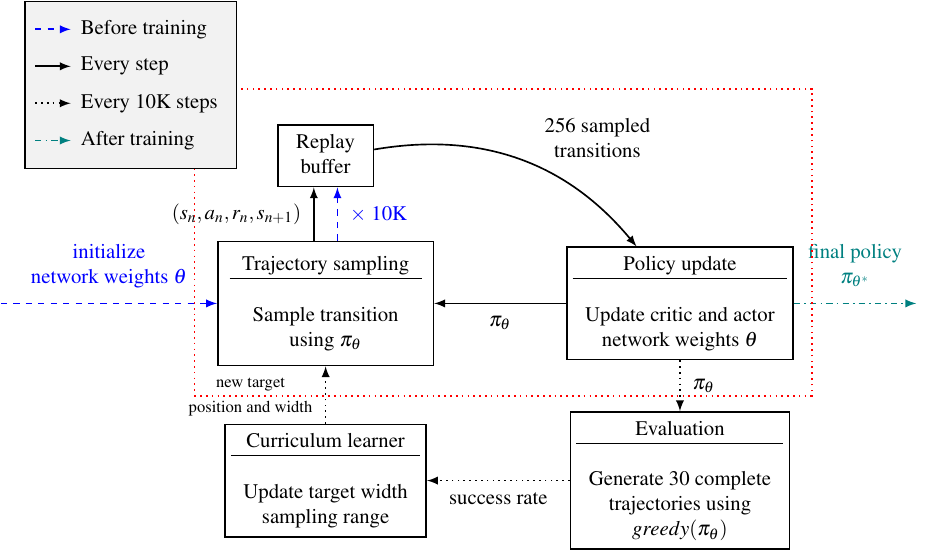}};
		\end{tikzpicture}
		\caption{\textbf{Reinforcement learning procedure.} Before training, the networks are initialized with random weights~$\theta$, and 10K transitions are generated using the resulting initial policy. These are stored in the replay buffer (blue dashed arrows). During training (red dotted box), trajectory sampling and policy update steps are executed alternately in each step. The targets used in the trajectory sampling part are generated by the curriculum learner, which is updated every 10K steps, based on an evaluation of the most recent (greedy) policy. As soon as the target width suggested by the curriculum learner falls below 1 cm, the training phase is completed and the final policy~$\pi_{\theta^*}$ is returned (teal dash-dotted arrow).}\label{fig:RL-training}
	\end{figure}

	An overview of the complete training procedure is given in Fig.~\ref{fig:RL-training}.

	\subsection*{Evaluation}
	For an evaluation of the trajectories resulting from the learned policy for different target conditions, we designed a discrete Fitts' Law type task.
	This task follows the ISO 9241-9 ergonomics standard and incorporates 13 equidistant targets arranged in a circle 50 cm in front of the body and placed 10 cm right of the right shoulder (Fig.~\ref{fig:fitts-law-type-task}).
	As soon as a target is reached and the end-effector remains inside for 100 ms, the next target is given to the learned policy. This also happens after 1.5 seconds, regardless of whether or not the episode was successful.
	
	Based on the recommendations from Guiard et al.\cite{Guiard09}, we determine different task difficulty conditions by sampling ``form and scale'', i.e., the \textit{Index of Difficulty (ID)} and the distance $D$ between the target centers are sampled independently, instead of using a distance-width grid.
	We use the \textit{Shannon Formulation}\cite{MacKenzie89} of Fitts' Law (equation~\eqref{eq:fitts-law-shannon}) to compute the resulting distance between the initial and target point $D$, given the target width $W$ and the ID:
	\begin{equation}
		\label{shannon-formulation}
		\text{ID} = \log_{2}\left(\frac{D}{W} + 1\right).
	\end{equation}
	The used combinations of distance, width, and ID can be found as Supplementary Table S1 online,
	and the resulting target setup is shown in Fig.~\ref{fig:fitts-law-type-task}a.
	
	The model executes 50 movements for each task condition and each direction, i.e., 6500 movements in total. All movements reached the target and remained inside for 100 ms within the given maximum movement time of 1.5 s.
	Plots for all task conditions and movement directions, together with their underlying data, can be found in a public repository\cite{Fischer20_zenodo}.
	
	In addition, an adaptive ``moving target'' mechanism is applied to generate elliptic movements from our learned policy.
	During training, the policy only learned to reach a given target as fast and accurate as possible -- it was never asked to follow a specific path accurately.
	For this reason, we make use of the following method.
	
	Initially, we place the first target on the ellipse such that $10\%$ of the complete curve needs to be covered clockwise within the first movement, starting at a fixed initial position (leftmost point on the ellipse). 
	In contrast to regular pointing tasks, the target already switches as soon as the movement (or rather the projection of the movement path onto the ellipse) covers more than half of this distance. The next target is then chosen so as to again create an incentive to cover the next $10\%$ of the elliptic curve.
	Thus, roughly 20 via-points in total are subsequently placed on the ellipse. 
	As shown in Fig.~\ref{fig:powerlaw}a, this indeed leads to fairly elliptic movements.
	
	For our evaluation, we use an ellipse with horizontal and vertical diameters of 15 cm and 6 cm (similar to the ellipse used by Harris and Wolpert\cite{HarrisWolpert98}), with its center placed 55 cm in front, 10 cm above, and 10 cm to the right of the shoulder.
	The task was performed for one minute, with end-effector position, velocity, and acceleration stored every 10 ms.
	
	Comprehensive data for all of these movements can also be found in a public repository\cite{Fischer20_zenodo}.

	\bibliography{bib}

\begin{thebibliography}{10}
\urlstyle{rm}
\expandafter\ifx\csname url\endcsname\relax
  \def\url#1{\texttt{#1}}\fi
\expandafter\ifx\csname urlprefix\endcsname\relax\def\urlprefix{URL }\fi
\expandafter\ifx\csname doiprefix\endcsname\relax\def\doiprefix{DOI: }\fi
\providecommand{\bibinfo}[2]{#2}
\providecommand{\eprint}[2][]{\url{#2}}

\bibitem{HarrisWolpert98}
\bibinfo{author}{Harris, C.~M.} \& \bibinfo{author}{Wolpert, D.~M.}
\newblock \bibinfo{journal}{\bibinfo{title}{Signal-dependent noise determines
  motor planning}}.
\newblock {\emph{\JournalTitle{Nature}}} \textbf{\bibinfo{volume}{394}},
  \bibinfo{pages}{780--4}, \doiprefix\url{10.1038/29528}
  (\bibinfo{year}{1998}).

\bibitem{Tanaka06}
\bibinfo{author}{Tanaka, H.}, \bibinfo{author}{Krakauer, J.~W.} \&
  \bibinfo{author}{Qian, N.}
\newblock \bibinfo{journal}{\bibinfo{title}{An optimization principle for
  determining movement duration}}.
\newblock {\emph{\JournalTitle{Journal of Neurophysiology}}}
  \textbf{\bibinfo{volume}{95}}, \bibinfo{pages}{3875--3886},
  \doiprefix\url{10.1152/jn.00751.2005} (\bibinfo{year}{2006}).
\newblock \bibinfo{note}{PMID: 16571740},
  \eprint{https://doi.org/10.1152/jn.00751.2005}.

\bibitem{Saul14}
\bibinfo{author}{Saul, K.~R.} \emph{et~al.}
\newblock \bibinfo{journal}{\bibinfo{title}{{Benchmarking of dynamic simulation
  predictions in two software platforms using an upper limb musculoskeletal
  model.}}}
\newblock {\emph{\JournalTitle{Computer methods in biomechanics and biomedical
  engineering}}} \textbf{\bibinfo{volume}{5842}}, \bibinfo{pages}{1--14},
  \doiprefix\url{10.1080/10255842.2014.916698} (\bibinfo{year}{2014}).

\bibitem{vanBeers03}
\bibinfo{author}{van Beers, R.~J.}, \bibinfo{author}{Haggard, P.} \&
  \bibinfo{author}{Wolpert, D.~M.}
\newblock \bibinfo{journal}{\bibinfo{title}{The role of execution noise in
  movement variability}}.
\newblock {\emph{\JournalTitle{Journal of Neurophysiology}}}
  \textbf{\bibinfo{volume}{91}}, \bibinfo{pages}{1050--1063},
  \doiprefix\url{10.1152/jn.00652.2003} (\bibinfo{year}{2004}).
\newblock \bibinfo{note}{PMID: 14561687},
  \eprint{https://doi.org/10.1152/jn.00652.2003}.

\bibitem{Sutton18}
\bibinfo{author}{Sutton, R.~S.} \& \bibinfo{author}{Barto, A.~G.}
\newblock \emph{\bibinfo{title}{Reinforcement Learning: An Introduction}}
  (\bibinfo{publisher}{A Bradford Book}, \bibinfo{address}{Cambridge, MA, USA},
  \bibinfo{year}{2018}).

\bibitem{Haarnoja18}
\bibinfo{author}{Haarnoja, T.} \emph{et~al.}
\newblock \bibinfo{title}{Soft actor-critic algorithms and applications}
  (\bibinfo{year}{2018}).
\newblock \eprint{1812.05905}.

\bibitem{Todorov12_mujoco}
\bibinfo{author}{{Todorov}, E.}, \bibinfo{author}{{Erez}, T.} \&
  \bibinfo{author}{{Tassa}, Y.}
\newblock \bibinfo{title}{Mujoco: A physics engine for model-based control}.
\newblock In \emph{\bibinfo{booktitle}{2012 IEEE/RSJ International Conference
  on Intelligent Robots and Systems}}, \bibinfo{pages}{5026--5033},
  \doiprefix\url{10.1109/IROS.2012.6386109} (\bibinfo{year}{2012}).

\bibitem{Flash85}
\bibinfo{author}{Flash, T.} \& \bibinfo{author}{Hogan, N.}
\newblock \bibinfo{journal}{\bibinfo{title}{The coordination of arm movements:
  an experimentally confirmed mathematical model}}.
\newblock {\emph{\JournalTitle{Journal of neuroscience}}}
  \textbf{\bibinfo{volume}{5}}, \bibinfo{pages}{1688--1703}
  (\bibinfo{year}{1985}).

\bibitem{HoffArbib93}
\bibinfo{author}{Hoff, B.} \& \bibinfo{author}{Arbib, M.~A.}
\newblock \bibinfo{journal}{\bibinfo{title}{Models of trajectory formation and
  temporal interaction of reach and grasp}}.
\newblock {\emph{\JournalTitle{Journal of Motor Behavior}}}
  \textbf{\bibinfo{volume}{25}}, \bibinfo{pages}{175--192},
  \doiprefix\url{10.1080/00222895.1993.9942048} (\bibinfo{year}{1993}).
\newblock \bibinfo{note}{PMID: 12581988},
  \eprint{https://doi.org/10.1080/00222895.1993.9942048}.

\bibitem{Nelson83}
\bibinfo{author}{Nelson, W.~L.}
\newblock \bibinfo{journal}{\bibinfo{title}{Physical principles for economies
  of skilled movements}}.
\newblock {\emph{\JournalTitle{Biol. Cybern.}}} \textbf{\bibinfo{volume}{46}},
  \bibinfo{pages}{135–147}, \doiprefix\url{10.1007/BF00339982}
  (\bibinfo{year}{1983}).

\bibitem{Artstein80}
\bibinfo{author}{Artstein, Z.}
\newblock \bibinfo{journal}{\bibinfo{title}{Discrete and continuous bang-bang
  and facial spaces or: look for the extreme points}}.
\newblock {\emph{\JournalTitle{Siam Review}}} \textbf{\bibinfo{volume}{22}},
  \bibinfo{pages}{172--185} (\bibinfo{year}{1980}).

\bibitem{Uno89}
\bibinfo{author}{Uno, Y.}, \bibinfo{author}{Kawato, M.} \&
  \bibinfo{author}{Suzuki, R.}
\newblock \bibinfo{journal}{\bibinfo{title}{Formation and control of optimal
  trajectory in human multijoint arm movement - minimum torque-change model}}.
\newblock {\emph{\JournalTitle{Biological cybernetics}}}
  \textbf{\bibinfo{volume}{61}}, \bibinfo{pages}{89--101},
  \doiprefix\url{10.1007/BF00204593} (\bibinfo{year}{1989}).

\bibitem{Todorov98_thesis}
\bibinfo{author}{Todorov, E.}
\newblock \bibinfo{title}{Studies of goal-directed movements}.
\newblock \bibinfo{howpublished}{Massachusetts Institute of Technology}
  (\bibinfo{year}{1998}).

\bibitem{Li04}
\bibinfo{author}{Li, W.} \& \bibinfo{author}{Todorov, E.}
\newblock \bibinfo{title}{Iterative linear quadratic regulator design for
  nonlinear biological movement systems.}
\newblock In \emph{\bibinfo{booktitle}{Proceedings of the 1st International
  Conference on Informatics in Control, Automation and Robotics, (ICINCO
  2004)}}, vol.~\bibinfo{volume}{1}, \bibinfo{pages}{222--229}
  (\bibinfo{year}{2004}).

\bibitem{Hoff94}
\bibinfo{author}{Hoff, B.}
\newblock \bibinfo{journal}{\bibinfo{title}{A model of duration in normal and
  perturbed reaching movement}}.
\newblock {\emph{\JournalTitle{Biol. Cybern.}}} \textbf{\bibinfo{volume}{71}},
  \bibinfo{pages}{481–488}, \doiprefix\url{10.1007/BF00198466}
  (\bibinfo{year}{1994}).

\bibitem{Bullock88}
\bibinfo{author}{Bullock, D.} \& \bibinfo{author}{Grossberg, S.}
\newblock \bibinfo{journal}{\bibinfo{title}{Neural dynamics of planned arm
  movements: emergent invariants and speed-accuracy properties during
  trajectory formation.}}
\newblock {\emph{\JournalTitle{Psychological review}}}
  \textbf{\bibinfo{volume}{95}}, \bibinfo{pages}{49} (\bibinfo{year}{1988}).

\bibitem{Plamondon98}
\bibinfo{author}{Plamondon, R.}
\newblock \bibinfo{journal}{\bibinfo{title}{A kinematic theory of rapid human
  movements: Part iii. kinetic outcomes}}.
\newblock {\emph{\JournalTitle{Biological Cybernetics}}}
  \textbf{\bibinfo{volume}{78}}, \bibinfo{pages}{133--145},
  \doiprefix\url{10.1007/s004220050420} (\bibinfo{year}{1998}).

\bibitem{Todorov02}
\bibinfo{author}{Todorov, E.} \& \bibinfo{author}{Jordan, M.~I.}
\newblock \bibinfo{journal}{\bibinfo{title}{Optimal feedback control as a
  theory of motor coordination}}.
\newblock {\emph{\JournalTitle{Nature Neuroscience}}}
  \textbf{\bibinfo{volume}{5}}, \bibinfo{pages}{1226--1235},
  \doiprefix\url{10.1038/nn963} (\bibinfo{year}{2002}).

\bibitem{Scott04}
\bibinfo{author}{Scott, S.}
\newblock \bibinfo{journal}{\bibinfo{title}{Optimal feedback control and the
  neural basis of volitional motor control}}.
\newblock {\emph{\JournalTitle{Nature reviews. Neuroscience}}}
  \textbf{\bibinfo{volume}{5}}, \bibinfo{pages}{532--46},
  \doiprefix\url{10.1038/nrn1427} (\bibinfo{year}{2004}).

\bibitem{Todorov04}
\bibinfo{author}{Todorov, E.}
\newblock \bibinfo{journal}{\bibinfo{title}{Optimality principles in
  sensorimotor control}}.
\newblock {\emph{\JournalTitle{Nature neuroscience}}}
  \textbf{\bibinfo{volume}{7}}, \bibinfo{pages}{907--15},
  \doiprefix\url{10.1038/nn1309} (\bibinfo{year}{2004}).

\bibitem{Shadmehr08}
\bibinfo{author}{Shadmehr, R.} \& \bibinfo{author}{Krakauer, J.}
\newblock \bibinfo{journal}{\bibinfo{title}{A computational neuroanatomy for
  motor control}}.
\newblock {\emph{\JournalTitle{Experimental Brain Research}}}
  \textbf{\bibinfo{volume}{185}}, \bibinfo{pages}{359--381}
  (\bibinfo{year}{2008}).

\bibitem{Diedrichsen10}
\bibinfo{author}{Diedrichsen, J.}, \bibinfo{author}{Shadmehr, R.} \&
  \bibinfo{author}{Ivry, R.~B.}
\newblock \bibinfo{journal}{\bibinfo{title}{The coordination of movement:
  optimal feedback control and beyond}}.
\newblock {\emph{\JournalTitle{Trends in Cognitive Sciences}}}
  \textbf{\bibinfo{volume}{14}}, \bibinfo{pages}{31 -- 39},
  \doiprefix\url{10.1016/j.tics.2009.11.004} (\bibinfo{year}{2010}).

\bibitem{Qian12}
\bibinfo{author}{Qian, N.}, \bibinfo{author}{Jiang, Y.},
  \bibinfo{author}{Jiang, Z.-P.} \& \bibinfo{author}{Mazzoni, P.}
\newblock \bibinfo{journal}{\bibinfo{title}{Movement duration, fitts's law, and
  an infinite-horizon optimal feedback control model for biological motor
  systems}}.
\newblock {\emph{\JournalTitle{Neural computation}}}
  \textbf{\bibinfo{volume}{25}}, \doiprefix\url{10.1162/NECO_a_00410}
  (\bibinfo{year}{2012}).

\bibitem{Shadmehr10}
\bibinfo{author}{Shadmehr, R.}, \bibinfo{author}{De~Xivry, J. J.~O.},
  \bibinfo{author}{Xu-Wilson, M.} \& \bibinfo{author}{Shih, T.-Y.}
\newblock \bibinfo{journal}{\bibinfo{title}{Temporal discounting of reward and
  the cost of time in motor control}}.
\newblock {\emph{\JournalTitle{Journal of Neuroscience}}}
  \textbf{\bibinfo{volume}{30}}, \bibinfo{pages}{10507--10516}
  (\bibinfo{year}{2010}).

\bibitem{Berret16}
\bibinfo{author}{Berret, B.} \& \bibinfo{author}{Jean, F.}
\newblock \bibinfo{journal}{\bibinfo{title}{Why don{\textquoteright}t we move
  slower? the value of time in the neural control of action}}.
\newblock {\emph{\JournalTitle{Journal of Neuroscience}}}
  \textbf{\bibinfo{volume}{36}}, \bibinfo{pages}{1056--1070},
  \doiprefix\url{10.1523/JNEUROSCI.1921-15.2016} (\bibinfo{year}{2016}).
\newblock \eprint{http://www.jneurosci.org/content/36/4/1056.full.pdf}.

\bibitem{Holzbaur05}
\bibinfo{author}{Holzbaur, K.~R.}, \bibinfo{author}{Murray, W.~M.} \&
  \bibinfo{author}{Delp, S.~L.}
\newblock \bibinfo{journal}{\bibinfo{title}{A model of the upper extremity for
  simulating musculoskeletal surgery and analyzing neuromuscular control}}.
\newblock {\emph{\JournalTitle{Annals of biomedical engineering}}}
  \textbf{\bibinfo{volume}{33}}, \bibinfo{pages}{829--840}
  (\bibinfo{year}{2005}).

\bibitem{Seth19}
\bibinfo{author}{Seth, A.}, \bibinfo{author}{Dong, M.},
  \bibinfo{author}{Matias, R.} \& \bibinfo{author}{Delp, S.}
\newblock \bibinfo{journal}{\bibinfo{title}{Muscle contributions to
  upper-extremity movement and work from a musculoskeletal model of the human
  shoulder}}.
\newblock {\emph{\JournalTitle{Frontiers in Neurorobotics}}}
  \textbf{\bibinfo{volume}{13}}, \bibinfo{pages}{90},
  \doiprefix\url{10.3389/fnbot.2019.00090} (\bibinfo{year}{2019}).

\bibitem{Delp07}
\bibinfo{author}{{Delp}, S.~L.} \emph{et~al.}
\newblock \bibinfo{journal}{\bibinfo{title}{Opensim: Open-source software to
  create and analyze dynamic simulations of movement}}.
\newblock {\emph{\JournalTitle{IEEE Transactions on Biomedical Engineering}}}
  \textbf{\bibinfo{volume}{54}}, \bibinfo{pages}{1940--1950}
  (\bibinfo{year}{2007}).

\bibitem{Seth18}
\bibinfo{author}{Seth, A.} \emph{et~al.}
\newblock \bibinfo{journal}{\bibinfo{title}{Opensim: Simulating musculoskeletal
  dynamics and neuromuscular control to study human and animal movement}}.
\newblock {\emph{\JournalTitle{PLOS Computational Biology}}}
  \textbf{\bibinfo{volume}{14}}, \bibinfo{pages}{1--20},
  \doiprefix\url{10.1371/journal.pcbi.1006223} (\bibinfo{year}{2018}).

\bibitem{Rosenbaum95}
\bibinfo{author}{Rosenbaum, D.~A.}, \bibinfo{author}{Loukopoulos, L.~D.},
  \bibinfo{author}{Meulenbroek, R.~G.}, \bibinfo{author}{Vaughan, J.} \&
  \bibinfo{author}{Engelbrecht, S.~E.}
\newblock \bibinfo{journal}{\bibinfo{title}{Planning reaches by evaluating
  stored postures}}.
\newblock {\emph{\JournalTitle{Psychological review}}}
  \textbf{\bibinfo{volume}{102}}, \bibinfo{pages}{28--67},
  \doiprefix\url{10.1037/0033-295x.102.1.28} (\bibinfo{year}{1995}).

\bibitem{Nakano99}
\bibinfo{author}{Nakano, E.} \emph{et~al.}
\newblock \bibinfo{journal}{\bibinfo{title}{Quantitative examinations of
  internal representations for arm trajectory planning: Minimum commanded
  torque change model}}.
\newblock {\emph{\JournalTitle{Journal of Neurophysiology}}}
  \textbf{\bibinfo{volume}{81}}, \bibinfo{pages}{2140--2155},
  \doiprefix\url{10.1152/jn.1999.81.5.2140} (\bibinfo{year}{1999}).
\newblock \bibinfo{note}{PMID: 10322055},
  \eprint{https://doi.org/10.1152/jn.1999.81.5.2140}.

\bibitem{Kawato93}
\bibinfo{author}{Kawato, M.}
\newblock \bibinfo{journal}{\bibinfo{title}{Optimization and learning in neural
  networks for formation and control of coordinated movement}}.
\newblock {\emph{\JournalTitle{Attention and Performance}}}
  \bibinfo{pages}{821--849} (\bibinfo{year}{1993}).

\bibitem{Kawato96}
\bibinfo{author}{Kawato, M.}
\newblock \bibinfo{journal}{\bibinfo{title}{Trajectory formation in arm
  movements: minimization principles and procedures}}.
\newblock {\emph{\JournalTitle{Advances in motor learning and control}}}
  \bibinfo{pages}{225--259} (\bibinfo{year}{1996}).

\bibitem{Cheema20}
\bibinfo{author}{Cheema, N.} \emph{et~al.}
\newblock \bibinfo{title}{Predicting mid-air interaction movements and fatigue
  using deep reinforcement learning}.
\newblock In \emph{\bibinfo{booktitle}{Proceedings of the 2020 CHI Conference
  on Human Factors in Computing Systems}}, CHI ’20, \bibinfo{pages}{1–13},
  \doiprefix\url{10.1145/3313831.3376701} (\bibinfo{publisher}{Association for
  Computing Machinery}, \bibinfo{address}{New York, NY, USA},
  \bibinfo{year}{2020}).

\bibitem{Berret11}
\bibinfo{author}{Berret, B.}, \bibinfo{author}{Chiovetto, E.},
  \bibinfo{author}{Nori, F.} \& \bibinfo{author}{Pozzo, T.}
\newblock \bibinfo{journal}{\bibinfo{title}{Evidence for composite cost
  functions in arm movement planning: An inverse optimal control approach}}.
\newblock {\emph{\JournalTitle{PLOS Computational Biology}}}
  \textbf{\bibinfo{volume}{7}}, \bibinfo{pages}{1--18},
  \doiprefix\url{10.1371/journal.pcbi.1002183} (\bibinfo{year}{2011}).

\bibitem{Lee19}
\bibinfo{author}{Lee, S.}, \bibinfo{author}{Park, M.}, \bibinfo{author}{Lee,
  K.} \& \bibinfo{author}{Lee, J.}
\newblock \bibinfo{journal}{\bibinfo{title}{Scalable muscle-actuated human
  simulation and control}}.
\newblock {\emph{\JournalTitle{ACM Trans. Graph.}}}
  \textbf{\bibinfo{volume}{38}}, \doiprefix\url{10.1145/3306346.3322972}
  (\bibinfo{year}{2019}).

\bibitem{Nakada18}
\bibinfo{author}{Nakada, M.}, \bibinfo{author}{Zhou, T.},
  \bibinfo{author}{Chen, H.}, \bibinfo{author}{Weiss, T.} \&
  \bibinfo{author}{Terzopoulos, D.}
\newblock \bibinfo{journal}{\bibinfo{title}{Deep learning of biomimetic
  sensorimotor control for biomechanical human animation}}.
\newblock {\emph{\JournalTitle{ACM Trans. Graph.}}}
  \textbf{\bibinfo{volume}{37}}, \doiprefix\url{10.1145/3197517.3201305}
  (\bibinfo{year}{2018}).

\bibitem{Si15}
\bibinfo{author}{Si, W.}, \bibinfo{author}{Lee, S.-H.},
  \bibinfo{author}{Sifakis, E.} \& \bibinfo{author}{Terzopoulos, D.}
\newblock \bibinfo{journal}{\bibinfo{title}{Realistic biomechanical simulation
  and control of human swimming}}.
\newblock {\emph{\JournalTitle{ACM Trans. Graph.}}}
  \textbf{\bibinfo{volume}{34}}, \doiprefix\url{10.1145/2626346}
  (\bibinfo{year}{2015}).

\bibitem{Fan20}
\bibinfo{author}{{Fan}, J.}, \bibinfo{author}{{Jin}, J.} \&
  \bibinfo{author}{{Wang}, Q.}
\newblock \bibinfo{title}{Humanoid muscle-skeleton robot arm design and control
  based on reinforcement learning}.
\newblock In \emph{\bibinfo{booktitle}{2020 15th IEEE Conference on Industrial
  Electronics and Applications (ICIEA)}}, \bibinfo{pages}{541--546},
  \doiprefix\url{10.1109/ICIEA48937.2020.9248350} (\bibinfo{year}{2020}).

\bibitem{Ben-itzhak2008}
\bibinfo{author}{Ben-Itzhak, S.} \& \bibinfo{author}{Karniel, A.}
\newblock \bibinfo{journal}{\bibinfo{title}{Minimum acceleration criterion with
  constraints implies bang-bang control as an underlying principle for optimal
  trajectories of arm reaching movements}}.
\newblock {\emph{\JournalTitle{Neural Computation}}}
  \textbf{\bibinfo{volume}{20}}, \bibinfo{pages}{779--812},
  \doiprefix\url{10.1162/neco.2007.12-05-077} (\bibinfo{year}{2008}).
\newblock \eprint{https://doi.org/10.1162/neco.2007.12-05-077}.

\bibitem{Berret2008}
\bibinfo{author}{Berret, B.} \emph{et~al.}
\newblock \bibinfo{journal}{\bibinfo{title}{The inactivation principle:
  Mathematical solutions minimizing the absolute work and biological
  implications for the planning of arm movements}}.
\newblock {\emph{\JournalTitle{PLOS Computational Biology}}}
  \textbf{\bibinfo{volume}{4}}, \bibinfo{pages}{1--25},
  \doiprefix\url{10.1371/journal.pcbi.1000194} (\bibinfo{year}{2008}).

\bibitem{Tieck18}
\bibinfo{author}{Tieck, J. C.~V.} \emph{et~al.}
\newblock \bibinfo{title}{Learning continuous muscle control for a multi-joint
  arm by extending proximal policy optimization with a liquid state machine}.
\newblock In \emph{\bibinfo{booktitle}{International Conference on Artificial
  Neural Networks}}, \bibinfo{pages}{211--221}
  (\bibinfo{organization}{Springer}, \bibinfo{year}{2018}).

\bibitem{Kidzinski18}
\bibinfo{author}{Kidzi{\'{n}}ski, {\L}.} \emph{et~al.}
\newblock \bibinfo{title}{Learning to run challenge solutions: Adapting
  reinforcement learning methods for neuromusculoskeletal environments}.
\newblock In \bibinfo{editor}{Escalera, S.} \& \bibinfo{editor}{Weimer, M.}
  (eds.) \emph{\bibinfo{booktitle}{The NIPS '17 Competition: Building
  Intelligent Systems}}, \bibinfo{pages}{121--153}
  (\bibinfo{publisher}{Springer International Publishing},
  \bibinfo{address}{Cham}, \bibinfo{year}{2018}).

\bibitem{Fitts54}
\bibinfo{author}{Fitts, P.~M.}
\newblock \bibinfo{journal}{\bibinfo{title}{The information capacity of the
  human motor system in controlling the amplitude of movement.}}
\newblock {\emph{\JournalTitle{Journal of experimental psychology}}}
  \textbf{\bibinfo{volume}{47}}, \bibinfo{pages}{381--391}
  (\bibinfo{year}{1954}).

\bibitem{MacKenzie89}
\bibinfo{author}{MacKenzie, I.~S.}
\newblock \bibinfo{journal}{\bibinfo{title}{A note on the information-theoretic
  basis for fitts’ law}}.
\newblock {\emph{\JournalTitle{Journal of Motor Behavior}}}
  \textbf{\bibinfo{volume}{21}}, \bibinfo{pages}{323--330},
  \doiprefix\url{10.1080/00222895.1989.10735486} (\bibinfo{year}{1989}).
\newblock \bibinfo{note}{PMID: 15136269},
  \eprint{https://doi.org/10.1080/00222895.1989.10735486}.

\bibitem{CrossmanGoodeve83}
\bibinfo{author}{Crossman, E. R. F.~W.} \& \bibinfo{author}{Goodeve, P.~J.}
\newblock \bibinfo{journal}{\bibinfo{title}{Feedback control of hand-movement
  and fitts' law}}.
\newblock {\emph{\JournalTitle{The Quarterly Journal of Experimental
  Psychology}}} \textbf{\bibinfo{volume}{35}}, \bibinfo{pages}{251--278}
  (\bibinfo{year}{1983}).

\bibitem{Meyer88}
\bibinfo{author}{Meyer, D.~E.}, \bibinfo{author}{Abrams, R.~A.},
  \bibinfo{author}{Kornblum, S.}, \bibinfo{author}{Wright, C.~E.} \&
  \bibinfo{author}{Keith~Smith, J.}
\newblock \bibinfo{journal}{\bibinfo{title}{Optimality in human motor
  performance: ideal control of rapid aimed movements.}}
\newblock {\emph{\JournalTitle{Psychological review}}}
  \textbf{\bibinfo{volume}{95}}, \bibinfo{pages}{340} (\bibinfo{year}{1988}).

\bibitem{Matthews96}
\bibinfo{author}{Matthews, P.}
\newblock \bibinfo{journal}{\bibinfo{title}{Relationship of firing intervals of
  human motor units to the trajectory of post‐spike after‐hyperpolarization
  and synaptic noise.}}
\newblock {\emph{\JournalTitle{The Journal of Physiology}}}
  \textbf{\bibinfo{volume}{492}} (\bibinfo{year}{1996}).

\bibitem{Takeda19}
\bibinfo{author}{Takeda, M.} \emph{et~al.}
\newblock \bibinfo{journal}{\bibinfo{title}{Explanation of fitts’ law in
  reaching movement based on human arm dynamics}}.
\newblock {\emph{\JournalTitle{Scientific Reports}}}
  \textbf{\bibinfo{volume}{9}}, \bibinfo{pages}{19804},
  \doiprefix\url{10.1038/s41598-019-56016-7} (\bibinfo{year}{2019}).

\bibitem{Lacquaniti83}
\bibinfo{author}{Lacquaniti, F.}, \bibinfo{author}{Terzuolo, C.} \&
  \bibinfo{author}{Viviani, P.}
\newblock \bibinfo{journal}{\bibinfo{title}{The law relating the kinematic and
  figural aspects of drawing movements}}.
\newblock {\emph{\JournalTitle{Acta Psychologica}}}
  \textbf{\bibinfo{volume}{54}}, \bibinfo{pages}{115 -- 130}
  (\bibinfo{year}{1983}).

\bibitem{Pollick97}
\bibinfo{author}{Pollick, F.~E.} \& \bibinfo{author}{Sapiro, G.}
\newblock \bibinfo{journal}{\bibinfo{title}{Constant affine velocity predicts
  the 13 power law of planar motion perception and generation}}.
\newblock {\emph{\JournalTitle{Vision Research}}}
  \textbf{\bibinfo{volume}{37}}, \bibinfo{pages}{347 -- 353},
  \doiprefix\url{https://doi.org/10.1016/S0042-6989(96)00116-2}
  (\bibinfo{year}{1997}).

\bibitem{VivianiSchneider91}
\bibinfo{author}{Viviani, P.} \& \bibinfo{author}{Schneider, R.}
\newblock \bibinfo{journal}{\bibinfo{title}{A developmental study of the
  relationship between geometry and kinematics in drawing movements.}}
\newblock {\emph{\JournalTitle{Journal of experimental psychology. Human
  perception and performance}}} \textbf{\bibinfo{volume}{17 1}},
  \bibinfo{pages}{198--218} (\bibinfo{year}{1991}).

\bibitem{deSperati97}
\bibinfo{author}{de'Sperati, C.} \& \bibinfo{author}{Viviani, P.}
\newblock \bibinfo{journal}{\bibinfo{title}{The relationship between curvature
  and velocity in two-dimensional smooth pursuit eye movements}}.
\newblock {\emph{\JournalTitle{The Journal of Neuroscience}}}
  \textbf{\bibinfo{volume}{17}}, \bibinfo{pages}{3932 -- 3945}
  (\bibinfo{year}{1997}).

\bibitem{VivianiMonoud90}
\bibinfo{author}{Viviani, P.} \& \bibinfo{author}{Mounoud, P.}
\newblock \bibinfo{journal}{\bibinfo{title}{Perceptuomotor compatibility in
  pursuit tracking of two-dimensional movements}}.
\newblock {\emph{\JournalTitle{Journal of Motor Behavior}}}
  \textbf{\bibinfo{volume}{22}}, \bibinfo{pages}{407--443},
  \doiprefix\url{10.1080/00222895.1990.10735521} (\bibinfo{year}{1990}).
\newblock \bibinfo{note}{PMID: 15117667},
  \eprint{https://doi.org/10.1080/00222895.1990.10735521}.

\bibitem{Viviani97}
\bibinfo{author}{Viviani, P.}, \bibinfo{author}{Baud-Bovy, G.} \&
  \bibinfo{author}{Redolfi, M.}
\newblock \bibinfo{journal}{\bibinfo{title}{Perceiving and tracking kinesthetic
  stimuli: further evidence of motor-perceptual interactions}}.
\newblock {\emph{\JournalTitle{Journal of experimental psychology. Human
  perception and performance}}} \textbf{\bibinfo{volume}{23}},
  \bibinfo{pages}{1232—1252}, \doiprefix\url{10.1037//0096-1523.23.4.1232}
  (\bibinfo{year}{1997}).

\bibitem{Hicheur05}
\bibinfo{author}{Hicheur, H.}, \bibinfo{author}{Vieilledent, S.},
  \bibinfo{author}{Richardson, M.}, \bibinfo{author}{Flash, T.} \&
  \bibinfo{author}{Berthoz, A.}
\newblock \bibinfo{journal}{\bibinfo{title}{Velocity and curvature in human
  locomotion along complex curved paths: a comparison with hand movements.}}
\newblock {\emph{\JournalTitle{Experimental brain research}}}
  \textbf{\bibinfo{volume}{162}}, \bibinfo{pages}{145--54},
  \doiprefix\url{10.1007/s00221-004-2122-8} (\bibinfo{year}{2005}).

\bibitem{Schaal01}
\bibinfo{author}{Schaal, S.} \& \bibinfo{author}{Sternad, D.}
\newblock \bibinfo{title}{Origins and violations of the 2/3 power law in
  rhythmic 3d movements}.
\newblock In \emph{\bibinfo{booktitle}{Experimental Brain Research}}, vol.
  \bibinfo{volume}{136}, \bibinfo{pages}{60--72} (\bibinfo{year}{2001}).
\newblock \bibinfo{note}{Clmc}.

\bibitem{Karklinsky15}
\bibinfo{author}{Karklinsky, M.} \& \bibinfo{author}{Flash, T.}
\newblock \bibinfo{journal}{\bibinfo{title}{Timing of continuous motor imagery:
  the two-thirds power law originates in trajectory planning}}.
\newblock {\emph{\JournalTitle{Journal of Neurophysiology}}}
  \textbf{\bibinfo{volume}{113}}, \bibinfo{pages}{2490--2499},
  \doiprefix\url{10.1152/jn.00421.2014} (\bibinfo{year}{2015}).
\newblock \bibinfo{note}{PMID: 25609105},
  \eprint{https://doi.org/10.1152/jn.00421.2014}.

\bibitem{Todorov98}
\bibinfo{author}{Todorov, E.}, \bibinfo{author}{Michael} \&
  \bibinfo{author}{Jordan, I.}
\newblock \bibinfo{journal}{\bibinfo{title}{Smoothness maximization along a
  predefined path accurately predicts the speed profiles of complex arm
  movements}}.
\newblock {\emph{\JournalTitle{Journal of Neurophysiology}}}
  \textbf{\bibinfo{volume}{80}}, \bibinfo{pages}{696--714}
  (\bibinfo{year}{1998}).

\bibitem{Gribble96}
\bibinfo{author}{Gribble, P.} \& \bibinfo{author}{Ostry, D.}
\newblock \bibinfo{journal}{\bibinfo{title}{Origins of the power law relation
  between movement velocity and curvature: modeling the effects of muscle
  mechanics and limb dynamics}}.
\newblock {\emph{\JournalTitle{Journal of neurophysiology}}}
  \textbf{\bibinfo{volume}{76}}, \bibinfo{pages}{2853—2860},
  \doiprefix\url{10.1152/jn.1996.76.5.2853} (\bibinfo{year}{1996}).

\bibitem{Schwartz94}
\bibinfo{author}{Schwartz, A.}
\newblock \bibinfo{journal}{\bibinfo{title}{Direct cortical representation of
  drawing}}.
\newblock {\emph{\JournalTitle{Science}}} \textbf{\bibinfo{volume}{265}},
  \bibinfo{pages}{540--542}, \doiprefix\url{10.1126/science.8036499}
  (\bibinfo{year}{1994}).
\newblock
  \eprint{https://science.sciencemag.org/content/265/5171/540.full.pdf}.

\bibitem{Flash07}
\bibinfo{author}{Flash, T.} \& \bibinfo{author}{Handzel, A.}
\newblock \bibinfo{journal}{\bibinfo{title}{Affine differential geometry
  analysis of human arm movements}}.
\newblock {\emph{\JournalTitle{Biological cybernetics}}}
  \textbf{\bibinfo{volume}{96}}, \bibinfo{pages}{577--601},
  \doiprefix\url{10.1007/s00422-007-0145-5} (\bibinfo{year}{2007}).

\bibitem{Morasso81}
\bibinfo{author}{Morasso, P.}
\newblock \bibinfo{journal}{\bibinfo{title}{Spatial control of arm movements}}.
\newblock {\emph{\JournalTitle{Experimental brain research}}}
  \textbf{\bibinfo{volume}{42}}, \bibinfo{pages}{223--227}
  (\bibinfo{year}{1981}).

\bibitem{Abend82}
\bibinfo{author}{Abend, W.}, \bibinfo{author}{Bizzi, E.} \&
  \bibinfo{author}{Morasso, P.}
\newblock \bibinfo{journal}{\bibinfo{title}{Human arm trajectory formation.}}
\newblock {\emph{\JournalTitle{Brain: a journal of neurology}}}
  \textbf{\bibinfo{volume}{105}}, \bibinfo{pages}{331--348}
  (\bibinfo{year}{1982}).

\bibitem{Cohen13}
\bibinfo{author}{Cohen, J.}
\newblock \emph{\bibinfo{title}{Statistical power analysis for the behavioral
  sciences}} (\bibinfo{publisher}{Academic press}, \bibinfo{year}{2013}).

\bibitem{Hausknecht15}
\bibinfo{author}{Hausknecht, M.} \& \bibinfo{author}{Stone, P.}
\newblock \bibinfo{journal}{\bibinfo{title}{Deep recurrent q-learning for
  partially observable mdps}}.
\newblock {\emph{\JournalTitle{arXiv preprint arXiv:1507.06527}}}
  (\bibinfo{year}{2015}).

\bibitem{Liu20}
\bibinfo{author}{Liu, J.}, \bibinfo{author}{Gu, X.} \& \bibinfo{author}{Liu,
  S.}
\newblock \bibinfo{title}{Reinforcement learning with world model}
  (\bibinfo{year}{2020}).
\newblock \eprint{1908.11494}.

\bibitem{Igl18}
\bibinfo{author}{Igl, M.}, \bibinfo{author}{Zintgraf, L.}, \bibinfo{author}{Le,
  T.~A.}, \bibinfo{author}{Wood, F.} \& \bibinfo{author}{Whiteson, S.}
\newblock \bibinfo{title}{Deep variational reinforcement learning for pomdps}.
\newblock In \emph{\bibinfo{booktitle}{International Conference on Machine
  Learning}}, \bibinfo{pages}{2117--2126} (\bibinfo{organization}{PMLR},
  \bibinfo{year}{2018}).

\bibitem{Jiang19}
\bibinfo{author}{Jiang, Y.}, \bibinfo{author}{Van~Wouwe, T.},
  \bibinfo{author}{De~Groote, F.} \& \bibinfo{author}{Liu, C.~K.}
\newblock \bibinfo{journal}{\bibinfo{title}{Synthesis of biologically realistic
  human motion using joint torque actuation}}.
\newblock {\emph{\JournalTitle{ACM Transactions on Graphics (TOG)}}}
  \textbf{\bibinfo{volume}{38}}, \bibinfo{pages}{1--12} (\bibinfo{year}{2019}).

\bibitem{Ikkala20}
\bibinfo{author}{Ikkala, A.} \& \bibinfo{author}{Hämäläinen, P.}
\newblock \bibinfo{title}{Converting biomechanical models from opensim to
  mujoco} (\bibinfo{year}{2020}).
\newblock \eprint{2006.10618}.

\bibitem{vanderHelm00}
\bibinfo{author}{van~der Helm, F. C.~T.} \& \bibinfo{author}{Rozendaal, L.~A.}
\newblock \emph{\bibinfo{title}{Musculoskeletal Systems with Intrinsic and
  Proprioceptive Feedback}}, chap.~\bibinfo{chapter}{11},
  \bibinfo{pages}{164--174} (\bibinfo{publisher}{Springer New York},
  \bibinfo{address}{New York, NY}, \bibinfo{year}{2000}).

\bibitem{Mnih16}
\bibinfo{author}{Mnih, V.} \emph{et~al.}
\newblock \bibinfo{title}{Asynchronous methods for deep reinforcement learning}
  (\bibinfo{year}{2016}).
\newblock \eprint{1602.01783}.

\bibitem{Eysenbach18}
\bibinfo{author}{Eysenbach, B.}, \bibinfo{author}{Gupta, A.},
  \bibinfo{author}{Ibarz, J.} \& \bibinfo{author}{Levine, S.}
\newblock \bibinfo{title}{Diversity is all you need: Learning skills without a
  reward function} (\bibinfo{year}{2018}).
\newblock \eprint{1802.06070}.

\bibitem{Ahmed19}
\bibinfo{author}{Ahmed, Z.}, \bibinfo{author}{Le~Roux, N.},
  \bibinfo{author}{Norouzi, M.} \& \bibinfo{author}{Schuurmans, D.}
\newblock \bibinfo{title}{Understanding the impact of entropy on policy
  optimization}.
\newblock In \bibinfo{editor}{Chaudhuri, K.} \& \bibinfo{editor}{Salakhutdinov,
  R.} (eds.) \emph{\bibinfo{booktitle}{Proceedings of the 36th International
  Conference on Machine Learning}}, vol.~\bibinfo{volume}{97} of
  \emph{\bibinfo{series}{Proceedings of Machine Learning Research}},
  \bibinfo{pages}{151--160} (\bibinfo{publisher}{PMLR}, \bibinfo{year}{2019}).

\bibitem{Fox15}
\bibinfo{author}{Fox, R.}, \bibinfo{author}{Pakman, A.} \&
  \bibinfo{author}{Tishby, N.}
\newblock \bibinfo{title}{Taming the noise in reinforcement learning via soft
  updates} (\bibinfo{year}{2015}).
\newblock \eprint{1512.08562}.

\bibitem{Vieillard21}
\bibinfo{author}{Vieillard, N.} \emph{et~al.}
\newblock \bibinfo{title}{Leverage the average: an analysis of kl
  regularization in rl} (\bibinfo{year}{2021}).
\newblock \eprint{2003.14089}.

\bibitem{Eysenbach19}
\bibinfo{author}{Eysenbach, B.} \& \bibinfo{author}{Levine, S.}
\newblock \bibinfo{title}{If maxent rl is the answer, what is the question?}
  (\bibinfo{year}{2019}).
\newblock \eprint{1910.01913}.

\bibitem{Eysenbach21}
\bibinfo{author}{Eysenbach, B.} \& \bibinfo{author}{Levine, S.}
\newblock \bibinfo{title}{Maximum entropy rl (provably) solves some robust rl
  problems} (\bibinfo{year}{2021}).

\bibitem{Haarnoja18b}
\bibinfo{author}{Haarnoja, T.}, \bibinfo{author}{Zhou, A.},
  \bibinfo{author}{Abbeel, P.} \& \bibinfo{author}{Levine, S.}
\newblock \bibinfo{title}{Soft actor-critic: Off-policy maximum entropy deep
  reinforcement learning with a stochastic actor}.
\newblock In \bibinfo{editor}{Dy, J.} \& \bibinfo{editor}{Krause, A.} (eds.)
  \emph{\bibinfo{booktitle}{Proceedings of the 35th International Conference on
  Machine Learning}}, vol.~\bibinfo{volume}{80} of
  \emph{\bibinfo{series}{Proceedings of Machine Learning Research}},
  \bibinfo{pages}{1861--1870} (\bibinfo{publisher}{PMLR},
  \bibinfo{year}{2018}).

\bibitem{Abdolmaleki18}
\bibinfo{author}{Abdolmaleki, A.} \emph{et~al.}
\newblock \bibinfo{title}{Maximum a posteriori policy optimisation}
  (\bibinfo{year}{2018}).
\newblock \eprint{1806.06920}.

\bibitem{Levine18}
\bibinfo{author}{Levine, S.}
\newblock \bibinfo{journal}{\bibinfo{title}{Reinforcement learning and control
  as probabilistic inference: Tutorial and review}}.
\newblock {\emph{\JournalTitle{CoRR}}}
  \textbf{\bibinfo{volume}{abs/1805.00909}} (\bibinfo{year}{2018}).
\newblock \eprint{1805.00909}.

\bibitem{Vulkan00}
\bibinfo{author}{Vulkan, N.}
\newblock \bibinfo{journal}{\bibinfo{title}{An economist’s perspective on
  probability matching}}.
\newblock {\emph{\JournalTitle{Journal of Economic Surveys}}}
  \textbf{\bibinfo{volume}{14}}, \bibinfo{pages}{101--118},
  \doiprefix\url{https://doi.org/10.1111/1467-6419.00106}
  (\bibinfo{year}{2000}).
\newblock
  \eprint{https://onlinelibrary.wiley.com/doi/pdf/10.1111/1467-6419.00106}.

\bibitem{Gruenwald04}
\bibinfo{author}{Grünwald, P.~D.} \& \bibinfo{author}{Dawid, A.~P.}
\newblock \bibinfo{journal}{\bibinfo{title}{{Game theory, maximum entropy,
  minimum discrepancy and robust Bayesian decision theory}}}.
\newblock {\emph{\JournalTitle{The Annals of Statistics}}}
  \textbf{\bibinfo{volume}{32}}, \bibinfo{pages}{1367 -- 1433},
  \doiprefix\url{10.1214/009053604000000553} (\bibinfo{year}{2004}).

\bibitem{Weir64}
\bibinfo{author}{Weir, M.~W.}
\newblock \bibinfo{journal}{\bibinfo{title}{Developmental changes in
  problem-solving strategies.}}
\newblock {\emph{\JournalTitle{Psychological review}}}
  \textbf{\bibinfo{volume}{71}}, \bibinfo{pages}{473} (\bibinfo{year}{1964}).

\bibitem{Gallistel90}
\bibinfo{author}{Gallistel, C.~R.}
\newblock \emph{\bibinfo{title}{The organization of learning.}}
  (\bibinfo{publisher}{The MIT Press}, \bibinfo{year}{1990}).

\bibitem{Peng18}
\bibinfo{author}{Peng, X.~B.}, \bibinfo{author}{Abbeel, P.},
  \bibinfo{author}{Levine, S.} \& \bibinfo{author}{van~de Panne, M.}
\newblock \bibinfo{journal}{\bibinfo{title}{Deepmimic: Example-guided deep
  reinforcement learning of physics-based character skills}}.
\newblock {\emph{\JournalTitle{ACM Transactions on Graphics}}}
  \textbf{\bibinfo{volume}{37}}, \bibinfo{pages}{1–14},
  \doiprefix\url{10.1145/3197517.3201311} (\bibinfo{year}{2018}).

\bibitem{Schulman17}
\bibinfo{author}{Schulman, J.}, \bibinfo{author}{Wolski, F.},
  \bibinfo{author}{Dhariwal, P.}, \bibinfo{author}{Radford, A.} \&
  \bibinfo{author}{Klimov, O.}
\newblock \bibinfo{title}{Proximal policy optimization algorithms}
  (\bibinfo{year}{2017}).
\newblock \eprint{1707.06347}.

\bibitem{Fujimoto18}
\bibinfo{author}{Fujimoto, S.}, \bibinfo{author}{van Hoof, H.} \&
  \bibinfo{author}{Meger, D.}
\newblock \bibinfo{title}{Addressing function approximation error in
  actor-critic methods} (\bibinfo{year}{2018}).
\newblock \eprint{1802.09477}.

\bibitem{Hasselt10}
\bibinfo{author}{Hasselt, H.~V.}
\newblock \bibinfo{title}{Double q-learning}.
\newblock In \bibinfo{editor}{Lafferty, J.~D.}, \bibinfo{editor}{Williams, C.
  K.~I.}, \bibinfo{editor}{Shawe-Taylor, J.}, \bibinfo{editor}{Zemel, R.~S.} \&
  \bibinfo{editor}{Culotta, A.} (eds.) \emph{\bibinfo{booktitle}{Advances in
  Neural Information Processing Systems 23}}, \bibinfo{pages}{2613--2621}
  (\bibinfo{publisher}{Curran Associates, Inc.}, \bibinfo{year}{2010}).

\bibitem{Kingma14}
\bibinfo{author}{Kingma, D.~P.} \& \bibinfo{author}{Ba, J.}
\newblock \bibinfo{title}{Adam: A method for stochastic optimization}
  (\bibinfo{year}{2014}).
\newblock \eprint{1412.6980}.

\bibitem{McCloskey89}
\bibinfo{author}{McCloskey, M.} \& \bibinfo{author}{Cohen, N.~J.}
\newblock \bibinfo{title}{Catastrophic interference in connectionist networks:
  The sequential learning problem}.
\newblock In \bibinfo{editor}{Bower, G.~H.} (ed.)
  \emph{\bibinfo{booktitle}{Psychology of Learning and Motivation}},
  vol.~\bibinfo{volume}{24}, \bibinfo{pages}{109 -- 165},
  \doiprefix\url{https://doi.org/10.1016/S0079-7421(08)60536-8}
  (\bibinfo{publisher}{Academic Press}, \bibinfo{year}{1989}).

\bibitem{Nair10}
\bibinfo{author}{Nair, V.} \& \bibinfo{author}{Hinton, G.~E.}
\newblock \bibinfo{title}{Rectified linear units improve restricted boltzmann
  machines}.
\newblock In \emph{\bibinfo{booktitle}{Proceedings of the 27th International
  Conference on International Conference on Machine Learning}}, ICML'10,
  \bibinfo{pages}{807–814} (\bibinfo{publisher}{Omnipress},
  \bibinfo{address}{Madison, WI, USA}, \bibinfo{year}{2010}).

\bibitem{TFAgents}
\bibinfo{author}{{Sergio Guadarrama, Anoop Korattikara, Oscar Ramirez, Pablo
  Castro, Ethan Holly, Sam Fishman, Ke Wang, Ekaterina Gonina, Neal Wu, Efi
  Kokiopoulou, Luciano Sbaiz, Jamie Smith, Gábor Bartók, Jesse Berent, Chris
  Harris, Vincent Vanhoucke, Eugene Brevdo}}.
\newblock \bibinfo{title}{{TF-Agents}: A library for reinforcement learning in
  tensorflow}.
\newblock \bibinfo{howpublished}{\url{https://github.com/tensorflow/agents}}
  (\bibinfo{year}{2018}).
\newblock \bibinfo{note}{[Online; accessed 25-June-2019]}.

\bibitem{Guiard09}
\bibinfo{author}{Guiard, Y.}
\newblock \bibinfo{title}{The problem of consistency in the design of fitts'
  law experiments: Consider either target distance and width or movement form
  and scale}.
\newblock In \emph{\bibinfo{booktitle}{Proceedings of the SIGCHI Conference on
  Human Factors in Computing Systems}}, CHI '09, \bibinfo{pages}{1809–1818},
  \doiprefix\url{10.1145/1518701.1518980} (\bibinfo{publisher}{Association for
  Computing Machinery}, \bibinfo{address}{New York, NY, USA},
  \bibinfo{year}{2009}).

\bibitem{Fischer20_zenodo}
\bibinfo{author}{Fischer, F.}, \bibinfo{author}{Bachinski, M.},
  \bibinfo{author}{Klar, M.}, \bibinfo{author}{Fleig, A.} \&
  \bibinfo{author}{Müller, J.}
\newblock \bibinfo{title}{Reinforcement learning control of a biomechanical
  model of the upper extremity (dataset)}.
\newblock \bibinfo{howpublished}{\emph{Zenodo}
  \url{https://dx.doi.org/10.5281/zenodo.4268230}} (\bibinfo{year}{2020}).

\end{thebibliography}

	\section*{Acknowledgements}
	
	We would like to thank Aldo A. Faisal (Imperial College London) for his very helpful advice and comments on the manuscript.
	
	\section*{Author contributions statement}
	
	F.F. and J.M. wrote the main manuscript text; M.B. adjusted the biomechanical model; F.F., M.B., and M.K. worked on the MuJoCo implementation and collected simulation data; F.F., M.K., and A.F. analyzed and validated the results; F.F. created the plots; F.F. conducted and J.M. supervised the research process.
	All authors reviewed the manuscript.
	
	\section*{Additional information}
	
	\textbf{Data Availability} The datasets generated during and/or analysed during the current study are available in a public repository, \url{https://doi.org/10.5281/zenodo.4268230}.
	
	~\\
	\noindent\textbf{Competing interests}
	The author(s) declare no competing interests.
	
	\begin{table}[!ht]
		\centering
		\begin{tabular}{|c|c|c|c|c|} 
			\hline
			\rule{0pt}{10pt}\noindent
			\multirow{2}{*}{Joint DOF} & \multicolumn{2}{c|}{Joint Angle Ranges (deg)} & \multicolumn{2}{c|}{Joint Torque Ranges (Nm)} \\
			\cline{2-5}
			\rule{0pt}{10pt}\noindent
			& Minimum & Maximum & Minimum & Maximum \\
			\hline
			\rule{0pt}{10pt}\noindent
			elevation angle & $-90$ & $130$ & $-36.01$ & $36.01$ \\
			\rule{0pt}{10pt}\noindent
			shoulder elevation & $0$ & $180$ & $-60.97$ & $60.97$ \\
			\rule{0pt}{10pt}\noindent
			shoulder rotation & $-90$ & $20$ & $-19.37$ & $19.37$ \\
			\rule{0pt}{10pt}\noindent
			elbow flexion & $0$ & $130$ & $-12.57$ & $12.57$ \\
			\rule{0pt}{10pt}\noindent
			pronation/supination & $-90$ & $90$ & $-1.03$ & $1.03$ \\
			\rule{0pt}{10pt}\noindent
			wrist deviation & $-10$ & $25$ & $-2.14$ & $2.14$ \\
			\rule{0pt}{10pt}\noindent
			wrist flexion & $-70$ & $70$ & $-1.53$ & $1.53$ \\
			\hline
		\end{tabular}
		\caption{\label{tab:joint_ranges}{\bf Joint ranges of individual DOFs.} Angle and torque ranges of all joint DOFs, which are actuated via second-order muscle dynamics (equation~\eqref{eq:activationmodel}). Moment arm scaling factors are defined as the magnitude of the torque range limits.}
	\end{table}

\end{document}